\documentclass{SciPost}
\hypersetup{
    colorlinks,
    linkcolor={red!50!black},
    citecolor={blue!50!black},
    urlcolor={blue!80!black}
}

\usepackage[bitstream-charter]{mathdesign}
\DeclareSymbolFont{usualmathcal}{OMS}{cmsy}{m}{n}
\DeclareSymbolFontAlphabet{\mathcal}{usualmathcal}

\usepackage{graphicx}
\usepackage{physics}
\usepackage{amsmath}

\usepackage{appendix}
\usepackage{bm}
\usepackage{dsfont}
\usepackage{float}
\usepackage{overpic}
\usepackage{tikz}
\usepackage[dvipsnames]{xcolor}
\usepackage[section]{placeins}
\usepackage{multirow}
\usepackage{nicefrac}
\usepackage[capitalize]{cleveref}
\usepackage{soul}

\definecolor{darkblue}{rgb}{0,0,.65}
\definecolor{darkgreen}{rgb}{0.28,0.41,0.19}
\definecolor{grey}{RGB}{180, 180, 180}
\definecolor{myorange}{HTML}{d9850f}

\providecommand{\imag}{\ensuremath{\mathrm{i}}}

\providecommand{\Ntrunc}{\ensuremath{\mathrm{N}_\mathrm{trunc}}}

\begin{document}

\begin{center}{\Large \textbf{
Basis dependence of Neural Quantum States for the Transverse Field Ising Model
}}\end{center}

\begin{center}
{
    Ronald Santiago Cortes\textsuperscript{1,2},
    Aravindh S. Shankar\textsuperscript{1},
    Marcello Dalmonte\textsuperscript{1,3},\\
    Roberto Verdel\textsuperscript{1},
    Nils Niggemann\textsuperscript{1}
}
\end{center}

\begin{center}
{\small
{\bf 1} The Abdus Salam International Center for Theoretical Physics (ICTP), Strada
Costiera 11, I-34151 Trieste, Italy
\\
{\bf 2} SISSA, Via Bonomea 265, I-34136 Trieste, Italy
\\
{\bf 3} Dipartimento di Fisica e Astronomia, Università di Bologna, via Irnerio 46, I-40126 Bologna, Italy
\\
}

\end{center}
 
\section*{Abstract}
\textbf{
Neural Quantum States (NQS) are powerful tools used to represent complex quantum many-body states in an increasingly wide range of applications. However, despite their popularity, at present only a rudimentary understanding of their limitations exists. In this work, we investigate the dependence of NQS on the choice of the computational basis, focusing on restricted Boltzmann machines.
Considering a family of rotated Hamiltonians corresponding to the paradigmatic transverse-field Ising model, we discuss the properties of ground states responsible for the dependence of NQS performance, namely the presence of ground state degeneracies as well as the uniformity of amplitudes and phases, carefully examining their interplay. 
We identify that the basis-dependence of the performance is linked to the convergence properties of a cluster or cumulant expansion of multi-spin operators---providing a framework to directly connect physical, basis-dependent properties, to performance itself. Our results provide insights that may be used to gauge the applicability of NQS to new problems and to identify the optimal basis for numerical computations.
}

\vspace{1em}
\tableofcontents

\section{Introduction}

Neural quantum states (NQS) have emerged as a powerful class of variational many-body wave functions, leveraging the expressive capabilities of modern neural-network architectures.

Despite the impressive numerical results achieved with these \emph{ans\"atze} in a multitude of models with interacting spins, fermions, or bosons~\cite{carleoSolvingQuantumManybody2017,PhysRevB.96.205152,PhysRevB.97.035116,chenEmpoweringDeepNeural2024,PhysRevLett.130.236401,viterittiTransformerWaveFunction2025,viterittiAccuracyRestrictedBoltzmann2022, chooTwodimensionalFrustratedJ1J22019, PhysRevB.97.195136, Lange2025,denisAccurateNeuralQuantum2025,machaczekNeuralQuantumState2025,chenNeuralNetworkaugmentedPfaffian2025, gu2025solving }, including their ability to efficiently represent highly entangled many-body states~\cite{DengQuantumEntanglementNeural2017,denisCommentCanNeural2025, PhysRevLett.134.076502}, several fundamental questions remain open regarding the factors that limit their performance.

Importantly, some notable statements have been made about the expressive power of given model classes~\cite{colluraDescriptivePowerNeuralNetworks2021,glasserNeuralnetworkQuantumStates2018 
,clarkUnifyingNeuralnetworkQuantum2018
,chenEquivalenceRestrictedBoltzmann2018,sharirNeuralTensorContractions2022,seeding_kaneko}, Jastrow and stabilizer states~\cite{peiCompactNeuralnetworkQuantum2021,glasserNeuralnetworkQuantumStates2018}. However, the existence of such an efficient representation does not automatically imply that such states can be learned efficiently.
For practical applications, one thus needs to address the question of whether a given neural network can not only \emph{efficiently encode}, but also \emph{efficiently learn} (through variational training) and \emph{efficiently sample from} the complex amplitudes of a quantum many-body wave function.

Although approaches improving sampling efficiency can be adapted from the extensive history of Markov chain Monte Carlo approaches, the question of learnability is more recent.
Among the first advances in this direction, the trainability was connected to geometric properties of the variational manifold, for example, through the quantum geometric tensor (QGT)~\cite{dashEfficiencyNeuralQuantum2024}, or through the geometry of the optimization landscape~\cite{parkGeometryLearningNeural2020}.

A central aspect of this fundamental question relates to how the sign structure of the wavefunction is learned and encoded.  Several works have proposed practical improvements~\cite{chenNeuralNetworkEvolution2022} or attempted clarification, for example, by designing an interpretable phase architecture and optimization protocol \cite{doschlImportanceCorrelationsNeural2025,szaboNeuralNetworkWave2020}, by analyzing sign structures through Fourier methods \cite{schurovLearningComplexityManybody2025}, and by studying how the non-stoquasticity of the Hamiltonian affects the learning process \cite{parkExpressivePowerComplexvalued2022}.
Other works have focused on ground state properties. For example, Ref.~\cite{borinApproximatingPowerMachinelearning2020} connects the RBM structure with a perturbation series. Recent advances in~\cite{yangWhenCanClassical2024} have proposed a more fundamental understanding of optimization complexity in terms of ground-state properties, particularly in the presence of long-range correlations or, more precisely, finite conditional mutual information. Further, Ref.~\cite{paul2025boundentanglementneuralquantum} has recently analyzed how the capacity of feed-forward NQS  to capture entanglement is bounded by the amount nonlinearities that are included in the architecture. Moreover, Ref.~\cite{doschlImportanceCorrelationsNeural2025} explores the basis dependence of correlations and provides explicit examples where capturing higher-order correlations is essential for an accurate representation of the quantum state.

Although some of these explanations are highly specific to the architecture and optimization method~\cite{dashEfficiencyNeuralQuantum2024,parkGeometryLearningNeural2020}, and others have focused solely on describing the sign structure ~\cite{doschlImportanceCorrelationsNeural2025,schurovLearningComplexityManybody2025,parkExpressivePowerComplexvalued2022}, a general characterization of when NQS can efficiently represent or learn a given many-body state with \emph{economical} resources such as a polynomial number of parameters or tractable training—remains an important open question.

In the following, we attempt to better understand this question by studying the effect of changing the computational basis. Since the wavefunction, but not its physical properties, is explicitly dependent on the basis on which the amplitudes are expressed, a basis transformation can dramatically modify the structural properties of the probability distribution associated with the wavefunction (through Born's rule), thus affecting representability and learnability in ways that are not yet fully understood~\cite{kovzic2025exploring,Sobral_2025}.

Notably, this basis-dependence is at odds with prevailing entanglement-based arguments for NQS performance, which are manifestly basis-independent.
To investigate this question, we employ a Restricted Boltzmann Machine (RBM). Our choice of neural network architecture is motivated by their comparatively simpler description, which makes them more suitable for studying fundamental questions regarding NQS.
We focus on the transverse-field Ising model and study how basis rotations affect its performance.
In order to separate learning and representability effects from limitations in Markov-chain sampling, we focus on small system sizes where all expectation values can be computed
by performing the summation over all basis states explicitly.
Based on our observations, we identify three key properties of the ground state wavefunction that determine the success or failure of RBM training: (i) the uniqueness of the ground state, (ii) the uniformity of wavefunction phases, and (iii) the uniformity of wavefunction amplitudes 
We further quantify the effects of non-uniformity by expanding the optimized RBM's in a cumulant series, studying the convergence properties of this expansion as a function of the basis rotation angle.

The remainder of this manuscript is organized as follows.
Section~\ref{NQS} introduces neural quantum states and the RBM architecture.
Section~\ref{sec:TFI-Model} describes the rotated transverse-field Ising Model and its physical properties.
Our numerical results and analysis are presented in Sect.~\ref{analysis}.
Subsequently, in Sect.~\ref{sec:cumulantExpansion}, we demonstrate that the accuracy of RBMs is determined by the accuracy of a truncated version of the cumulant expansion.
Finally, conclusions and perspectives for future work are discussed in Sect.~\ref{conclusions}.

\section{Neural Quantum States }
\label{NQS}

Given a $L$ spin-$1/2$ Hilbert space $\mathcal{H}^{(L)}=\mathbb{C}^{2^L}$ and a Hamiltonian of the system of interest $H$, a generic wave function describing the state of the quantum system can be expanded in a complete basis set $\{\ket{s}\}$ as
\begin{equation}
    \ket{\Psi}=\sum_{s}\Psi(s)\ket{s}, 
\end{equation}
where $\Psi(s)=\langle s|\Psi \rangle$ can be a complicated function of the binary vector $s$. Note that, typically, one chooses the basis of tensor products of single spin states along the reference quantization axis, that is, $|s\rangle \equiv |\sigma^z_1 \sigma^z_2 \cdots \sigma^z_N\rangle$, with $\sigma^z_i =\pm 1$ being the eigenvalue of the Pauli-$z$ operator. This canonical choice is often referred to as the ``computational basis''. 

While the solution to the quantum many-body problem requires determining the exponentially many wave function amplitudes, $\Psi(s)$, the main idea of the seminal work by Carleo and Troyer~\cite{carleoSolvingQuantumManybody2017} is the realization that such amplitudes can be compactly parametrized and generated in terms of artificial neural networks, that is a function which maps the input configuration $s$ to a complex number:
\begin{equation}
    \Psi(s) \approx \Psi_\omega(s). 
\end{equation}
Here, the neural network depends on a set of learnable parameters $\omega$. In the context of ground-state search problems, these parameters are optimized in an iterative procedure, according to the variational principle~\cite{beccaQuantumMonteCarlo2017}.
This principle implies that minimization of the variational energy of a state $\ket{\Psi_\omega}$ leads to an approximation of the ground state of the Hamiltonian $H$: 
\begin{align}
    E_\omega &= \frac{\langle \Psi_{\omega} |H| \Psi_{\omega}\rangle}{\langle \Psi_{\omega} | \Psi_{\omega}\rangle} = \frac{\sum_s |\Psi_\omega(s)|^2 E^\textrm{loc}_\omega(s)}{\sum_s |\Psi_\omega(s)|^2} \label{eq:energy} \\
    &E^\textrm{loc}_\omega(s) \equiv \frac{\langle{s|H|\Psi_{\omega}\rangle}}{\langle{s|\Psi_{\omega}\rangle}} = \sum_{s'} H_{ss'} \frac{\Psi_\omega(s')}{\Psi_\omega(s)}, \label{eq:Eloc} 
\end{align}
where $E^\textrm{loc}_\omega(s)$ is the \emph{local energy}. 

Note that in practical problems, the above summations must usually be stochastically estimated via Markov chain Monte Carlo (MCMC)~\cite{beccaQuantumMonteCarlo2017}.
In this work, we restrict ourselves to small system sizes, where we may perform the summation over the entire Hilbert space.

\subsection{Restricted Bolzmann machines as NQS}
While several architectures have been considered in NQS studies (see Ref.~\cite{langeArchitecturesApplicationsReview2024} for a recent review), in this work, we focus on a family of models that has enjoyed considerable popularity due to their simplicity and efficient evaluation (see, e.g., Refs.~\cite{carleoSolvingQuantumManybody2017, nomuraRestrictedBoltzmannMachine2017, DengQuantumEntanglementNeural2017, melkoRestrictedBoltzmannMachines2019, parkGeometryLearningNeural2020, parkExpressivePowerComplexvalued2022, colluraDescriptivePowerNeuralNetworks2021, triguerosSimplicityMeanfieldTheories2024, dashEfficiencyNeuralQuantum2024}), namely, \emph{restricted Boltzmann machines} (RBMs). The RBM architecture leads to the following expression for the variational wave function:
\begin{equation}
\label{eq:RBM}
  \Psi_{\mathrm{RBM}}(s)=\exp{\sum_{i=1}^L a_i s_i }\times \prod_{j=1}^{M} \cosh \left( \sum_{i}s_iW_{ij}+b_j \right),
\end{equation}
where $a_1, a_2, \dots, a_L$ and $b_1, b_2, \dots, b_M$ are called visible and hidden biases, respectively, and $W_{ij}$ are the elements of a matrix of weights $(L\times M)$, which connects visible and hidden units. Thus, the RBM \emph{ansatz} in Eq.~\eqref{eq:RBM} has a total of $L+ M + (L\times M)$ parameters.  According to the representability theorems, if the number of hidden units $M$ is large enough, RBMs become universal approximators of any discrete probability distribution over the visible variables~\cite{lerouxRepresentationalPowerRestricted2008}.  Note that in order to allow for the RBM ansatz to represent arbitrary wave functions, we let the variational parameters be complex valued. An alternative strategy is to represent the amplitudes and phases with an individual neural network, which can be advantageous in cases of a severe sign problem \cite{szaboNeuralNetworkWave2020,bukovLearningGroundState2021,chenNeuralNetworkEvolution2022}

In the following, we will use the shorthand notation $\Psi_\omega(s) \equiv \braket{s}{\Psi_\omega} $ to refer specifically (unless otherwise stated) to the RBM wave function ansatz in Eq.~\eqref{eq:RBM}, with $\omega\equiv (a_i, b_j, W_{ij})$. Furthermore, the density of hidden units $\alpha\equiv M/N$ allows one to control the number of variational parameters included in the RBM.

\subsection{Stochastic reconfiguration}
The algorithm used here to train the RBMs is the \emph{Stochastic Reconfiguration} (SR) method~\cite{beccaQuantumMonteCarlo2017}, which we briefly summarize below. At each step $n$, the energy from \cref{eq:Eloc} is estimated for the state $\Psi_{\omega}$. The gradient of this energy can be obtained efficiently if the derivatives of the neural network $\partial_{\omega_k} \Psi_\omega$ are known (or can be computed via automatic differentiation). Updating the variational parameters in the direction of this gradient towards low energies, one may thus find progressively better approximations to the ground state until convergence. This approach is known as stochastic gradient descent.

As an improvement to this scheme, the SR method takes into account the geometry of the variational manifold defined by the parameters of  the neural network  $\omega$.
In this method, the parameters are updated as 
\begin{equation}
    \omega_k^{'}=\omega_k+\Delta \omega_k, \quad   \Delta \omega_k=\eta \sum_{k'}(S + \varepsilon \mathds{1})_{kk'}^{-1}f_{k'}, \label{eq:SR_update}
\end{equation}
where $\eta>0$ is the so-called learning rate and is small enough to guarantee convergence, $\epsilon>0$ a small diagonal shift used to guarantee the existence of an inverse\footnote{Note that for practical purposes, it is usually best to avoid the construction of the matrix $S$ instead obtain $\Delta \omega_k$ by an iterative solution of the linear system \cref{eq:SR_update}.}. Generalized forces are defined as $f_k=\langle E^\textrm{loc}_\omega O^\star_k \rangle - \langle E^\textrm{loc}_\omega\rangle \langle O^\star_k \rangle$, where $\expval{\dots}$ refers to expectation values with respect to the quantum state $\Psi_\omega$. In principle, the matrix $S$ can be any positive definite matrix; however, the SR prescription defines the matrix:
\begin{equation}
    S_{k,k'}= \langle O_k^\star O_{k'} \rangle-\langle O_k^\star \rangle \langle O_{k'} \rangle,
\end{equation}
where $O_k(x)=\partial_{\omega_k}\ln{\braket{s}{\Psi_\omega}}$. For pure quantum states, the matrix $S$ is also called the \emph{quantum Fisher matrix}~\cite{parkGeometryLearningNeural2020}. For the particular case of RBMs, the operators $O_k$ can be computed as \cite{carleoSolvingQuantumManybody2017}:
\begin{equation}
\langle x|O_{k}|x'\rangle=O_k(x)\delta_{x,x'}=\delta_{x,x'}\times\left\{
    \begin{matrix}
        s_i, && &k=a_i \\
        \tanh{\chi_j(s)}, && &k=b_j \\
        s_i \tanh{\chi_j(s)}, && &k=W_{ij}       
    \end{matrix}
    \right.
\end{equation}
with $\chi_j(s)=b_j+\sum_i W_{ij}s_i$.

\section{Rotated transverse field Ising model}
\label{sec:TFI-Model}

\subsection{Rotated Hamiltonian representation}

The one-dimensional (1D) transverse-field Ising model describes a system of $L$ spin-$1/2$ in an external field, placed on a linear lattice:
\begin{equation}
\label{eq:rot-TFIM}
    H(\theta)=-\sum_{i=1}^{L-1}\tilde{\sigma}_i^z(\theta) \tilde{\sigma}_{i+1}^z(\theta)-g\sum_{i=1}^L \tilde{\sigma}_{i}^x(\theta),
\end{equation}
with the rotated Pauli matrices:
\begin{equation}
\label{eq:rotations}
\tilde{\sigma}_i^x(\theta)=\sigma^x_i\cos{\theta}+\sigma^z_i\sin{\theta}, \quad 
    \tilde{\sigma}^z_i(\theta)=\sigma^z_i\cos{\theta}-\sigma^x_i\sin{\theta}.
\end{equation}
Any such rotation leaves the model's  eigenspectrum and its physical properties invariant. However, the representation of the ground state wave function, in the computational basis (eigenbasis of $\sigma^z$), including its sign structure, depends explicitly on the angle $\theta$.
At any rotation, a conserved quantity is given as the global parity
\begin{equation}
    P=\prod_{i=1}^L \tilde{\sigma}_i^x(\theta).
\end{equation}
At $\theta=0$, we recover the ``conventional'' form $H(\theta=0)=-\sum_{i=1}^{L-1}{\sigma}_i^z {\sigma}_{i+1}^z-g\sum_{i=1}^L {\sigma}_{i}^x,$  whereas $\theta=\pi/2$ amounts to exchanging $\sigma_i^z \leftrightarrow \sigma_i^x$.

\subsection{Phase diagram and ground state}
\begin{figure}
    \centering
    \includegraphics[width=1.0\linewidth]{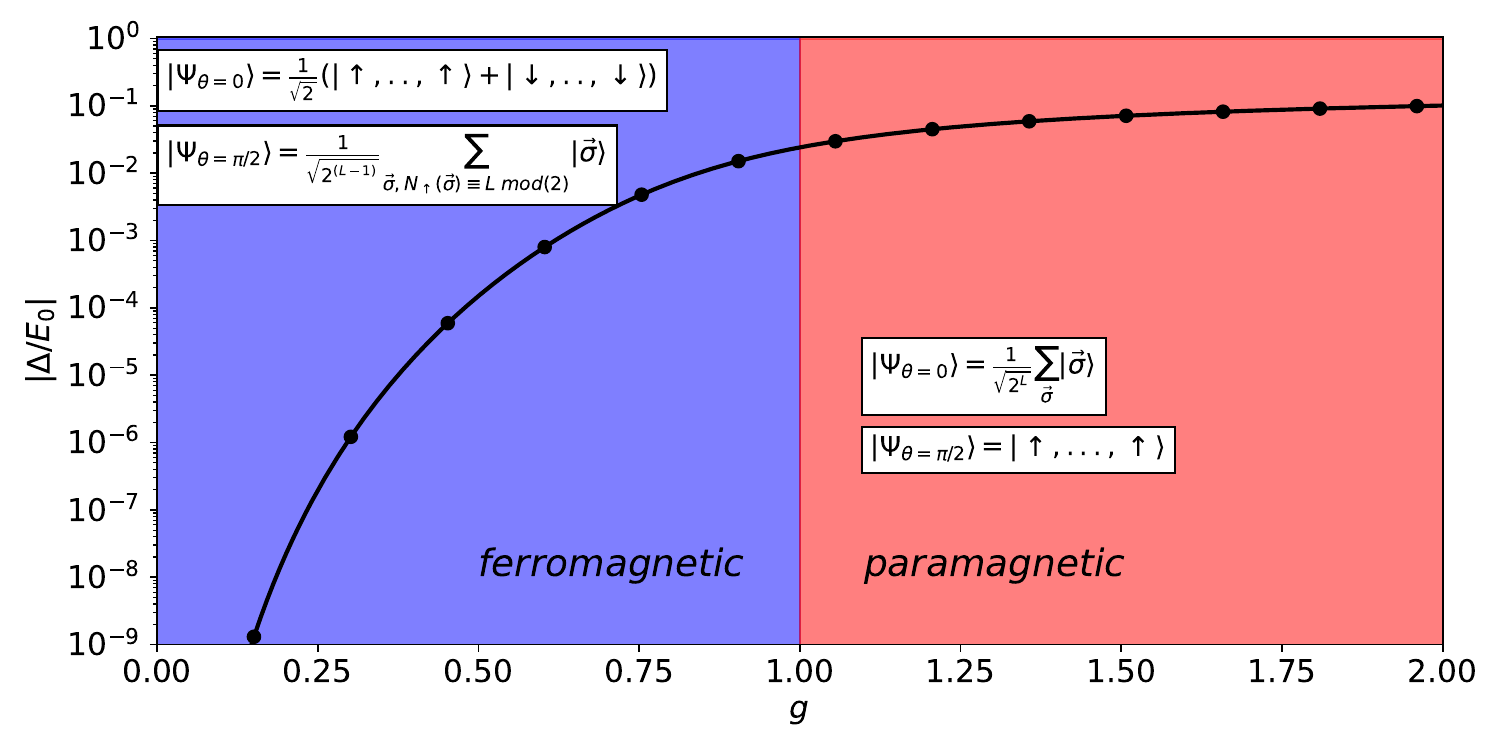}
    \caption{Energy gap ratio for a finite-size TFIM  with $L=10$, obtained via exact diagonalization of \cref{eq:rot-TFIM}. The blue part of the phase diagram corresponds to the symmetry-broken phase, whereas the red side corresponds to the paramagnetic phase. These two phases are separated by a second-order quantum phase transition at $g=1$. (This description of the phase diagram of the TFIM refers to its thermodynamic limit.) Also shown is the representation of the ground state in the computational $\sigma_z$ basis with $\theta = 0, \pi/2$  in the limits $g \to 0$ and $g \to \infty$.}
    \label{fig:phase_diagram}
\end{figure}

The phase diagram of the 1D TFIM, is shown in \cref{fig:phase_diagram} (for more details, see, e.g., Ref.~\cite{sachdevQuantumPhaseTransitions2011}).
\paragraph{Symmetry-broken phase} 
For $|g|<1$, the model features a $\mathbb{Z}_2$ symmetry-broken phase, characterized by long-range ferromagnetic order. In the thermodynamic limit, the ground-state manifold is two-fold degenerate, and at $g=0$, is spanned by the states $|\cdots \uparrow \uparrow \uparrow \cdots \rangle_{\tilde z}$ and $|\cdots \downarrow \downarrow \downarrow \cdots\rangle_{\tilde z}$, where $\ket{\dots}_{\tilde z}$ are eigenstates of the rotated operator $\tilde{\sigma}^z$. As shown in \cref{fig:phase_diagram}, its representation in the computational basis ranges from very simple, to relatively complex: For instance at $\theta = \pi/2$, this state is represented as a superposition of all basis states with even parity $P$.
The symmetry-broken phase is gapped, i.e. the energy difference between the ground state and the lowest excited eigenstates remains finite in the thermodynamic limit.
However, for finite systems, the energy splitting between the lowest eigenstates is typically of the order $Je^{-L/\xi}$, with $\xi\sim-1/\ln (|g|)$~\cite{mbeng2024quantum,huSpontaneousSymmetryBreaking2023}, i.e., it is exponentially small in system size.

\paragraph{Paramagnetic phase}
For $|g|>1$, the system in the disordered or paramagnetic phase, its ground state is separated from excitations by a gap  and preserves the spin-flip symmetry, but it is non-degenerate. The ground state is adiabatically connected to the limit $g \to \infty$ where it becomes $|\cdots \rightarrow \rightarrow\rightarrow \cdots\rangle_{\Tilde{z}}$. At $\theta=\pi/2$, this state has the particularly simple form of an eigenstate in the computational basis $\ket{\uparrow,\uparrow,\dots}_z$.

\paragraph{Quantum critical point}
At $|g|=1$, the model undergoes a quantum phase transition. In the thermodynamic limit it is characterized by gapless excitations and its low-energy physics is described by the two-dimensional Ising conformal field theory with central charge $c=1/2$ \cite{calabreseEntanglementEntropyConformal2009, sachdevQuantumPhaseTransitions2011}. 

\section{Key factors for NQS performance}
\label{analysis}
The performance of a NQS's ability to express a ground state can be measured in several ways. Most rigorously, the infidelity 
\begin{equation}
    I =1-\left\vert\frac{\braket{\Psi_0}{\Psi_{\omega}}}{\braket{\Psi_{\omega}}{\Psi_{\omega}}}\right\vert^2
\end{equation}
expresses the true distance of the state with respect to the ground state. Here, we quantify it using the more practical energy error 
\begin{equation}
    \Delta E = E_\omega - E_0.
\end{equation}
We note that for states close to the ground state  $\ket{\Psi_\omega} \approx \sqrt{1-I} \ket{\Psi_0} + \sqrt{I} \ket{\Psi_1}$ the two quantities are related via the spectral gap $\Delta$ as
\begin{equation}
    \Delta E  \approx \Delta \times  I. 
\end{equation}

We now turn to studying the possible links between properties of the Hamiltonian and the performance

We identify three main aspects that make a ground state amenable to be efficiently approximated by NQS: (i) The uniqueness of the ground state, i.e. whether there is a unique, gapped and non-degenerate ground state, (ii) the phase uniformity, and (iii) the amplitude uniformity. 

\begin{figure}
\centering 
\includegraphics[width=0.8\linewidth]{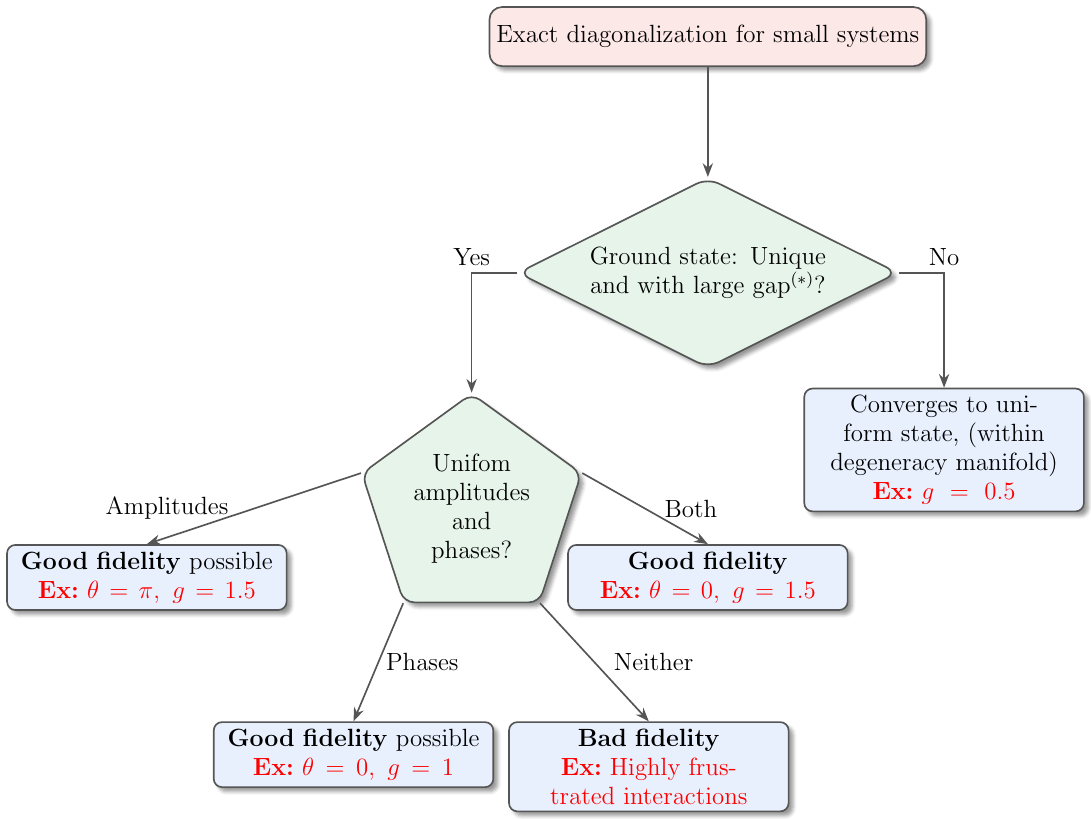}

\caption{Flowchart summarizing whether an RBM is expected to accurately learn the ground state of a given Hamiltonian.
$^{(*)}$Relevant is the size of the spectral gap $\Delta $ for a \emph{finite} system, in relation to optimization parameters such as the learning rate, or more precisely, the ratio $\frac{\Delta }{\eta f_k}$~\cite{beccaQuantumMonteCarlo2017}.}

\label{fig:flowchart} 
\end{figure}

Shown in Fig.~\ref{fig:flowchart} is a condensed flow-chart summarizing our findings, which are detailed in the following. In particular, we highlight that the performance properties are not fully decided by these aspects individually but rather by the intricate interplay between them. 

\subsection{Ground state uniqueness}
We find the uniqueness of the ground state to be a decisive factor in the outcome of the RBM's training. Importantly, we make the following observation:

\begin{center}
\noindent
\setlength{\fboxsep}{10pt}
\colorbox{grey!20}{
    \parbox{0.9\linewidth}{
        \textbf{If the true ground state has a (near) degeneracy, then the RBM will converge to the simplest possible superposition of degenerate states.}
    }
}
\end{center}

In the above, a ''simple'' superposition of states is one which has all positive signs in the computational basis, and whose amplitudes are nearly uniform.

We first discuss the simplicity of the sign structure. To characterize it, we compute the sign average, defined as: 
\begin{equation}
\expval{\textrm{sign}(\Psi)} = \sum_s |\Psi(s)|^2 \mathrm{sgn}(\Psi(s))
\label{eq:4.1.sign_av}
\end{equation}

\begin{figure}
    \centering
    \includegraphics[width=0.95\linewidth]{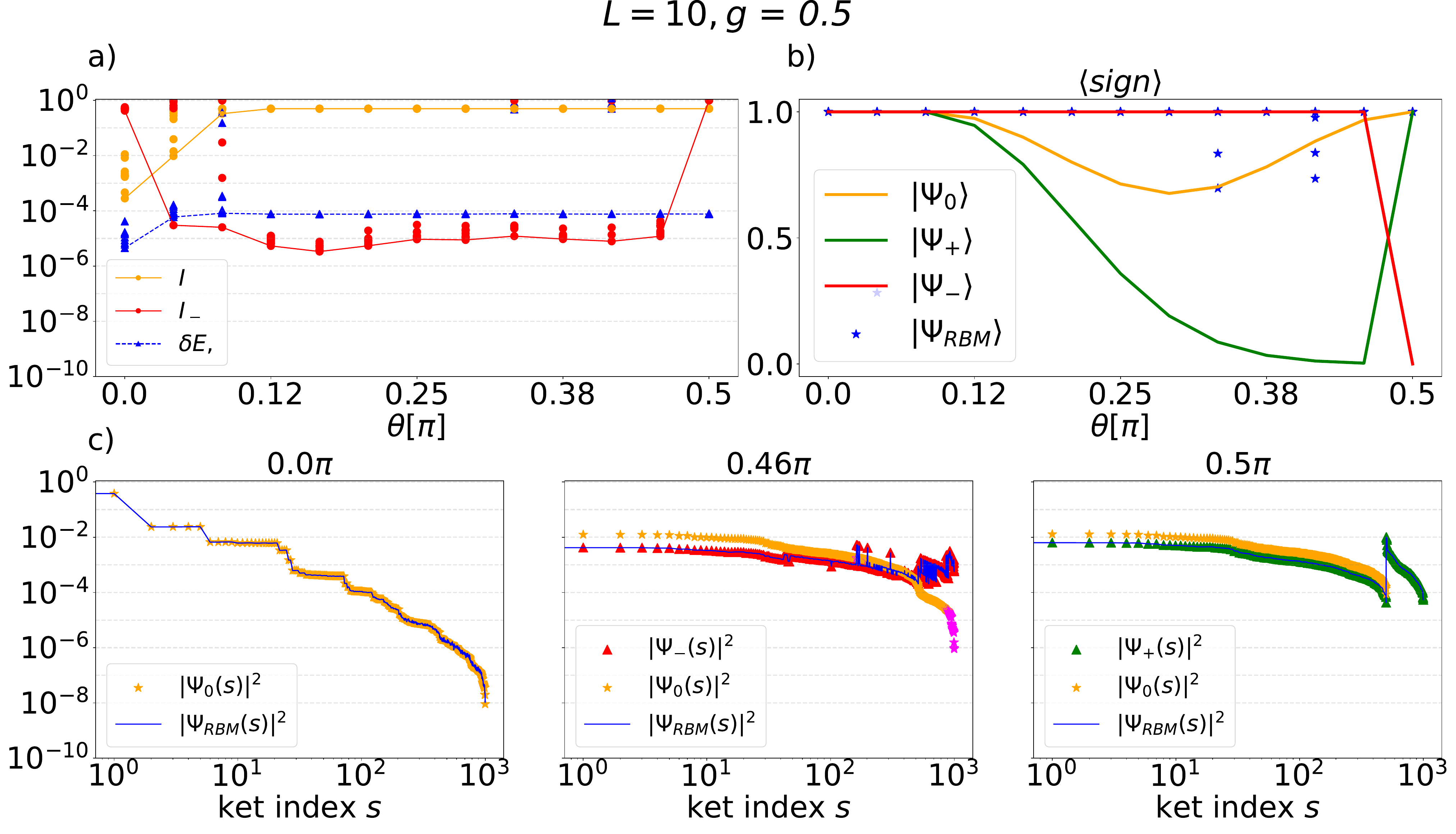}
    \caption{a) Relative energy error $\delta E$ and infidelities of the RBM wavefunction respect to $\ket{\Psi_0}$ and $\ket{\Psi_-}$ obtained by energy minimization over different basis rotations. For each angle, 10 independent realizations are represented by the markers. The dotted line connects the realizations with the lowest energy at each angle, and the solid line shows the infidelity of these selected realizations for $L=10$ and $g=0.5$. b)
    Average sign of the true ground state $|\Psi_0\rangle$, the superpositions $|\Psi_{\pm}\rangle =\frac{1}{\sqrt{2}}(|\Psi_0\rangle \pm |\Psi_1\rangle)$ ($\Psi_1$ indicates the first excited state)
    and 10 independent RBM realizations. c) Sorted probabilities for the angles $\theta=0,\;0.46\pi$ and $0.5\pi$. The color indicates the sign of the wavefunction components: orange points correspond to positive amplitudes of $\Psi$, while magenta points correspond to negative ones. The RBM amplitudes are plotted as a blue line and $\Psi_{\pm}$ in green and red respectively. The RBM calculations here were done with $\alpha=1$.}
    \label{fig:Lambda05Overview}
\end{figure}

An example that supports the previous statement can be found at $g=0.5$, which features two nearly degenerate states distinguished by the eigenvalue of their parity. We refer to the ground state as $\ket{\Psi}$  and to the excited (odd) parity state as $\ket{\Psi_1}$.

For $\theta=0,\pi/2$, the Hamiltonian is stoquastic. Therefore, the ground state can be written as a vector with purely positive elements. As shown in \cref{fig:Lambda05Overview}a), in $\theta=0$, $\ket{\Psi}$ is learned by the RBM with good accuracy. For $0<\theta<\pi/2$, the ground state acquires a sign structure in the computational basis, leading to a worse variational energy and a large infidelity. As shown in panel b), the average sign of the RBM remains positive, according to the superposition $\ket{\Psi_-} = \ket{\Psi} - \ket{\Psi_1}$. We note that while we allow for the RBM to have arbitrary complex amplitudes, the wave function is found to be real up to numerical precision up to a global phase.

Finally, even if the sign structure of the ground state is unproblematic, the amplitudes can still pose a challenge: An example is given by the case $\theta=\pi/2$ in the ferromagnetic phase. In this case, the spins are aligned along the $X$-axis. In the computational basis, this corresponds to a superposition of all states with parity $\prod_i \sigma^z_i = 1$. In order to faithfully represent this state, the RBM needs to correctly project out all odd-parity states, which is a difficult task, complicated even further by the small gap between the ground and the first excited state. As a result, the RBM instead converges to the symmetric superposition $\ket{\Psi_+} = \ket{\Psi} + \ket{\Psi_1}$, which has a simpler amplitude structure in the computational basis. 

We emphasize that the absence of a unique ground state only implies poor performance if the true ground state has either a complicated sign or amplitude structure.
Previous work on the sign problem for NQS show similar results for frustated antiferromagnets \cite{szaboNeuralNetworkWave2020}.
In other cases, such as for $\theta = 0$, the true ground state is easily learned by the RBM despite the near degeneracy.
 
\subsection{Phase and amplitude uniformity}
\label{subsec:phaseAmpUniformity}
A well-known challenge for NQS is the representation of wavefunctions with complex sign structures \cite{bukovLearningGroundState2021,ouImprovingNeuralNetwork2025,szaboNeuralNetworkWave2020}. Here, we find that it is indeed typically easier to learn ground states that have a uniform phase structure on the chosen basis. In contrast, the uniformity of the amplitudes is often overlooked, which also plays an important role. 
To explain the complicated interplay between these two aspects, we consider the optimization landscape, which is determined by the matrix elements of the generally complex Hamiltonian. To illustrate the different effects of amplitudes and phases on the variational energy landscape, we write the Hamiltonian matrix elements as
\begin{align}
    H_{s's} &\equiv -|H_{s's}| e^{\imag \Theta_{s,s'}}
\end{align}
The variational energy expectation value for a state $\Psi_\omega(s) = A_\omega(s)e^{i \phi_\omega(s)} $ is then given by \cref{eq:Eloc} as 
\begin{align}
    E_\omega &= -\frac{1}{\sum_s A_\omega(s)^2} \times \sum_{s s'} |H_{s's}| A_\omega(s'){A_\omega(s)} e^{\imag (\varphi_\omega(s')-\varphi_\omega(s))+\Theta_H(s,s')}.
\end{align}
The optimal energy depends on a complicated interplay between amplitudes and phases. This difficulty can be illuminated by considering two limiting cases.\\

\paragraph{Stoquastic Hamiltonians:}

Stoquastic Hamiltonians are defined as $H_{s\neq s'}\leq 0$. As this is equivalent to $\Theta_H(s,s') = 0$, the variational energy is minimized by a state with $\varphi(s) = 0$, independent of $A_\omega(s)$. This phase structure can be easily found by optimization procedures, as decreasing the imaginary part of the wavefunction always results in a lower energy.
These considerations hold for Hamiltonians that are related to stoquastic ones via a local gauge transformation, i.e. $\Theta_H(x,x') = \vartheta_H(x')-\vartheta_H(x)$.  Again, in this case, the problem of optimizing amplitudes and phases decouples, as the optimal phase structure can be found by fitting $\varphi_\omega(s) = \vartheta_H(s)$.\\

In both cases, the energy functional contains only terms related to the amplitudes of the wavefunction over the states:

\begin{equation}
    E^\textrm{stoch}_\omega = - \frac{1}{\sum_s A_\omega(s)^2} \times \sum_{s s'} |H_{s's}| A_\omega(s'){A_\omega(s)}. \label{eq:stoch_H_energy}
\end{equation}

\begin{figure}
    \centering
    \includegraphics[width=1.0\linewidth]{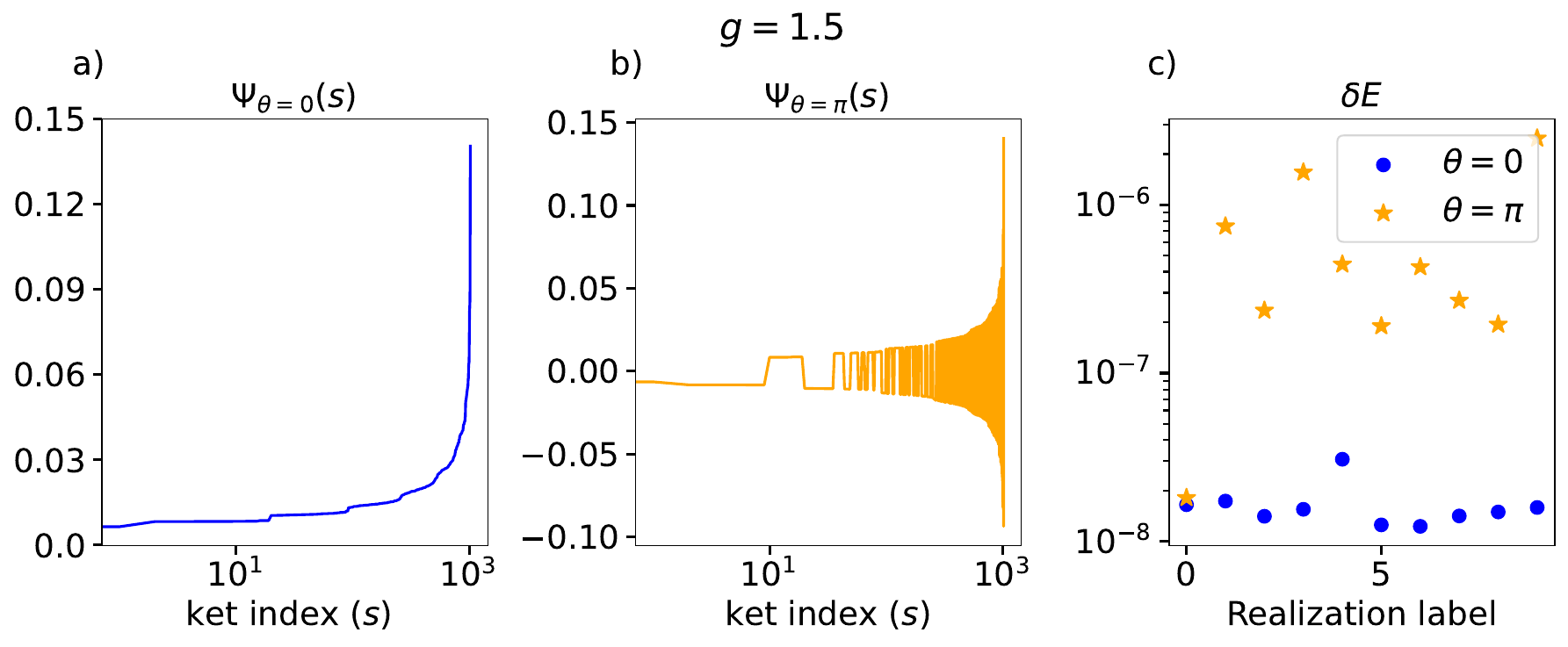}
    \caption{Amplitudes  $\Psi_{\theta}(s)= \langle{s|\psi_\textrm{RBM}\rangle_{\theta}}$ of the ground state wavefunction of the TFIM with $g=1.5$ for a) $\theta = 0$ and b) $\theta = \pi$, sorted by increasing absolute value $|\Psi(s)|$. c) Energy error for different realizations of the optimization.}
    \label{fig:Comparisonlambda15}
\end{figure}

In general, global rotations do not act as gauge transformations. However, in the context of the quantum Ising model, a gauge transformation can be constructed by rotating around either the x- or y-axis at an angle of $\pm \pi$.
An interesting consequence of this occurs in the paramagnetic phase, where the ground state exhibits a (+1) parity. Because of this global $Z_2$ symmetry, the wavefunction is symmetric under a global spin flip: $\Psi(s)=\Psi(-s)$. Apply a global $\pi$ rotation around the y-axis to the state, it can be expressed as:
\begin{equation}
    \exp\left\{i\frac{\pi}{2}\sum_j \sigma_j^y\right\}\Psi(s)\ket{s}=\exp\left\{  i\pi N_{\uparrow}(s)\right\}\Psi(s)\ket{-s}
\end{equation}
By introducing a change of variables $s'=-s$ we obtain $\exp\left\{  i\pi N_{\downarrow}(s')\right\}\Psi(s')\ket{s'}$. Consequently, the rotation results in the addition of a local phase (see \cref{fig:Comparisonlambda15}b). This sign rule is denominated in literature as the Marshall sign rule. Even though the Hamiltonian has become non-stoquastic (the $-X$ term is mapped to $X$), the additional phase 
can be cured by a local transformation. However, as illustrated in panel \cref{fig:Comparisonlambda15}c) this change of basis leads to an increased energy error across nearly all realizations. 
Crucially, the Marshall sign can be incorporated simply by adjusting the visible biases of the RBM (see Ref.~\cite{peiCompactNeuralnetworkQuantum2021} and Appendix \ref{sec:pi_rotation_RBM}). This suggests that representing the non-stoquastic wavefunction is inherently as efficient as the stoquastic case. Consequently, the observed rise in energy error is associated to optimization difficulty and cannot be related to a representational limitation.

However, even in the stoquastic cases, the performance of RBMs remains clearly basis dependent and can be poor if the amplitudes are complicated. 
\paragraph{Uniform amplitudes: \label{sec_uniform_amplitudes}} 
We may also assume that the ground state amplitudes are uniform $|\Psi(s)|^2 = 1/2^L$, such that the energy reduces to
\begin{equation}
    E_\omega = - \frac{1}{2^L}\sum_{s s'} |H_{s's}| e^{\imag (\varphi_\omega(s')-\varphi_\omega(s))+\Theta_H(s,s')}. \label{eq:E_uniform_amplitudes}
\end{equation}

Minimizing $E_\omega$ thus corresponds to minimizing the phase differences $\Theta_H(s,s') - (\varphi_\omega(s')-\varphi_\omega(s))$ particularly for those pairs $s,s'$ where $|H_{s,s'}|$ is large.

Clearly, this second scenario is rather artificial in nature, as ground states typically do not have uniform amplitudes. However, if the amplitudes are nearly uniform, or more generally, when the distribution is fairly well spread over all possible spin configurations,\footnote{This property of a quantum state is, in some contexts, referred to as \emph{anticoncentration}~\cite{PRXQuantum.3.010333}.} 
this leads to an equivalent problem as \cref{eq:E_uniform_amplitudes} however, with a renormalized Hamiltonian amplitude
\begin{equation}
    |H_{ss'}| \rightarrow |\tilde{H}_{ss'}|= |H_{ss'}| A_\omega(s') A_\omega(s)/\sum_s{A_\omega(s)^2}
    \label{eq:uniform_amp}
\end{equation} 

Previous studies \cite{szaboNeuralNetworkWave2020} have proposed a decoupled ansatz in which the amplitudes and phases are parameterized by separate neural networks. This architecture facilitates a targeted learning protocol where the amplitudes are held constant while only the phases are optimized. Such a regime closely mirrors our preceding analysis, resulting in an optimization landscape governed by Equation \ref{eq:uniform_amp}.  This optimization strategy has been successfully applied to frustrated antiferromagnets, as it is generally understood that the sign structure in these systems depends only weakly on the amplitude profile \cite{szaboNeuralNetworkWave2020}.

\paragraph{General case:}
If neither phases nor amplitudes take a simple form, the optimization does not decouple, leading to a complicated interplay between the two: For given configurations $s$ and $s'$ given by nonzero elements of $H_{ss'}$, the energy may be lowered by either decreasing the product of the amplitudes $A_\omega(s)A_\omega(s')$, or by adjusting the phases $\varphi_\omega(s), \varphi_\omega(s')$ to match $\Theta_H(s,s')$, in which case the energy may be lowered by choosing $A_\omega(s)A_\omega(s')$ to be large. This can be expected to lead to a generally complicated energy manifold where local minima can pose challenges for minimization~\cite{ouImprovingNeuralNetwork2025}.

\begin{figure}
    \centering
    \includegraphics[width=0.9\linewidth]{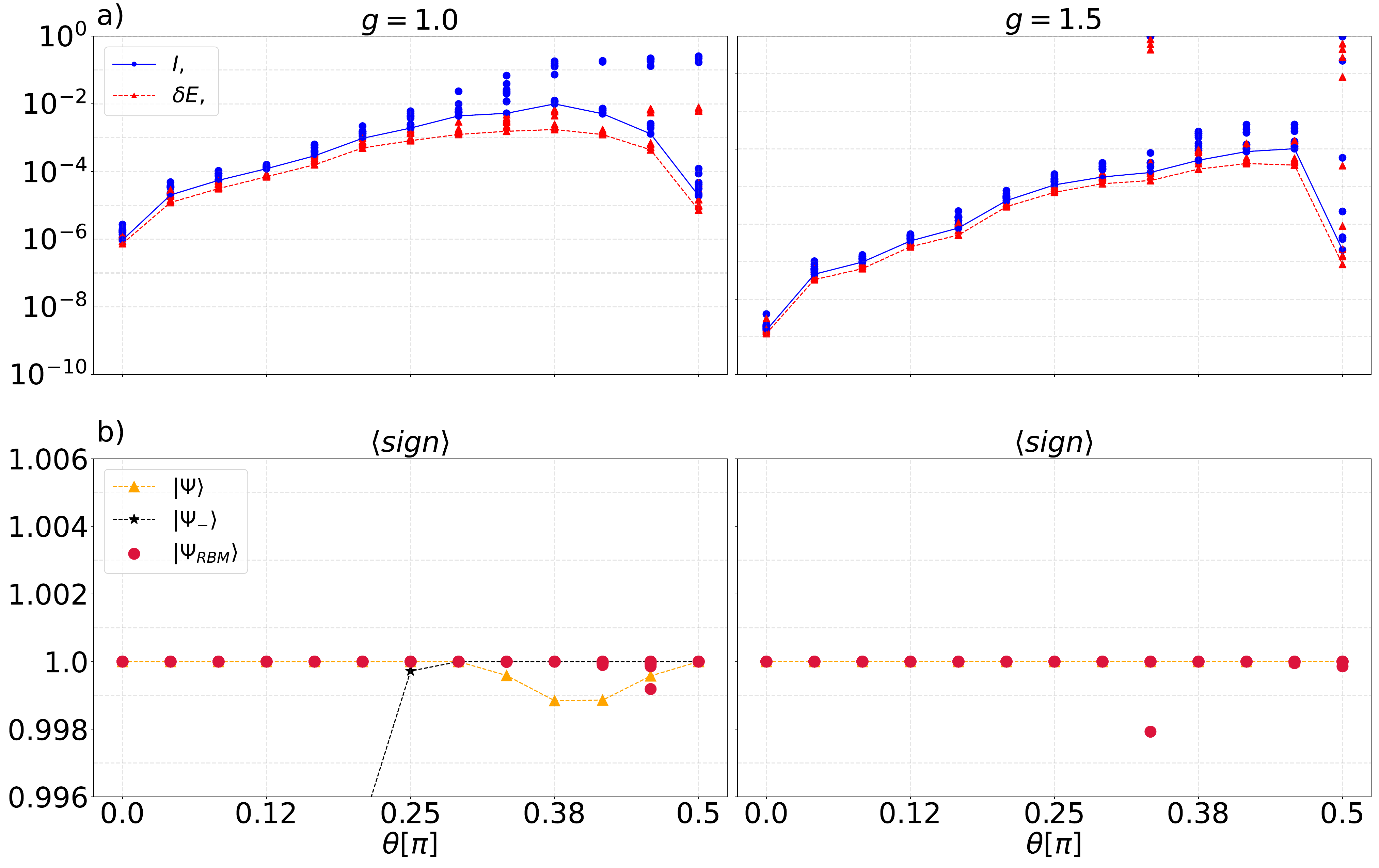}
    \caption{a) Relative energy error (red) and infidelity (blue) of the RBM wavefunction as a function of the rotation angle $\theta$ for different values of the transverse field strength $g$. Point correspond to  independent RBM realizations. The best-performing realizations are connected by a line. b) Average sign of the ground state wavefunction in the computational basis.}
    \label{fig:Lam1_15Comparison}
\end{figure}

In \Cref{fig:Lam1_15Comparison}a), we show the relative energy error and infidelity of the RBM at $g=1$ and $g=1.5$, respectively, as a function of $\theta$. While the TFIM Hamiltonian is stoquastic only at $\theta=0$ and $\theta=\pi/2$, as visible from panel b) the average sign of the ground state is not affected at $g=1.5$ and is only mildly affected at $g=1.0$ by the basis transformation. Nonetheless, the infidelity of the RBM increases by several orders of magnitude for angles above $\theta=0$, before finally lowering slightly towards $\theta=\pi/2$.
As ground state properties and the sign structure are unaffected, the only explanation for this behavior must be given by the distribution of wave-function amplitudes. As detailed in \cref{sec:TFI-Model}, for $\theta=0, g \rightarrow \infty$, the ground state is close to a uniform superposition, obtained by an RBM with $a_i = W_{ij} = b_j = 0$, whereas it drastically becomes narrower towards increasing $\theta$, finally approaching a singular state $\ket{\uparrow,\uparrow \dots}_z$ at $\theta=\pi/2$, which can be represented by $a_i \rightarrow \infty$. 
It can thus be concluded that a good first indicator of the RBM's performance is given by the degree of uniformity of the wave function amplitudes. In \cref{sec:cumulantExpansion}, we will further quantify this statement. 

We note that previous works have identified a declined NQS performance for peaked wavefunctions, which has been associated with issues in the Monte Carlo sampling, see, for example, Ref.~\cite{malyshev2024neural}. Importantly, our observations indicate that, even when not relying on sampling, peaked wavefunctions are generally more difficult to efficiently learn by RBMs.

\section{Convergence analysis based on a truncated cumulant expansion}

\label{sec:cumulantExpansion}
While it was argued above that a good qualitative understanding of RBM performance can be obtained from the uniformity of phases and amplitudes, i.e. the deviation of the probability distribution from a uniform one, this statement does not indicate how severe the impact of a given ``non-uniformity''  will be.
We generalize this notion of non-uniformity of a distribution, by considering the influence of higher correlations which can be analyzed via the \emph{truncated cumulant expansion} (or \emph{ coupled cluster expansion}) of an arbitrary state $\Phi(s)$ \cite{kuboGeneralizedCumulantExpansion1962,bartlettCoupledclusterTheoryQuantum2007}
\begin{align}
\label{eq:clusterexp}
\Phi(s) &= \exp\!\left[
  \log(\mathcal{N})+ \sum_{i} c_i s_i
  + \sum_{i<j} c_{ij} s_i s_j
  + \sum_{i<j<k} c_{ijk} s_i s_j s_k
  + \cdots
\right]\\
&\equiv \exp\left[\sum_{A \subseteq \mathcal{P}(\{1,\dots,L\})} c_{A}(\Phi) S_A(s) \right],
\end{align}
where each coefficient $c_{A}$ is associated with a subset $A \subseteq \mathcal{P}(\{1,\dots,L\})$ and the terms
$S_A = \prod_{i\in A} s_i$ are monomials of binary variables. The notation $\mathcal{P}(S)$ denotes the power set of $S$.
This expansion can be thought of as a generalization of the well-known Jastrow wave function ansatz~\cite{jastrowManybodyProblemStrong1955}, to higher correlation orders. 
Moreover, \cref{eq:clusterexp} is related to the Hadamard-Walsh transformation, recently employed in the context of neural quantum states in Refs.~\cite{schurovLearningComplexityManybody2025,doschlImportanceCorrelationsNeural2025} as detailed in Appendix~\ref{sec:Hadamard}. At each order $n$, the number of coefficients is given as $\binom{L}{n}$, yielding a total number of $2^L$ terms. 

To allow for a tunable comparison to other variational wavefunctions, we define the \emph{truncated cumulant expansion} $\Phi^{(\Ntrunc)}(s)$ to degree $\Ntrunc$ as follows:
\begin{itemize}
\item[1.] Compute \emph{all} $2^L$ coefficients $c_{A}(\Psi)$ of the \emph{exact} ground state $\Psi(s)$ in \cref{eq:clusterexp}.
\item[2.] Find the set $\{A\}_\textrm{largest}$ corresponding to the indices of the $\Ntrunc$ \emph{by magnitude} largest  coefficients $|c_{A}(\Psi)|$.
\item[3.] Define $\Phi^{(\Ntrunc)}(s) \equiv \exp(\sum_{A \in \{A\}_\textrm{largest}}  c_A(\Phi) S_A(s))$
\end{itemize}
Put more simply, $\Phi^{(N)}(s)$ is obtained by retaining only the $N$ most significant coefficients in the cumulant expansion of the ground state $\Psi(s) = \Psi^{(2^L)}(s)$. We emphasize the difference of this approach from a more direct truncation scheme where all coefficients up to a certain order $n$ are retained~\cite{schmittQuantumDynamicsTransversefield2018, PhysRevB.103.165103}.
At finite truncation level, the different convergence properties between the cumulant and the Hadamard-Welsh expansion employed in Refs.~\cite{schurovLearningComplexityManybody2025,doschlImportanceCorrelationsNeural2025} become apparent: Mapping the cumulant expansion to the corresponding Hadamard-Walsh expansion by expanding the \emph{exponent} in \cref{eq:clusterexp} will in general generate many high-order terms. For instance, a Jastrow function
\begin{equation}
    \exp(\sum_{i<j} c_{ij} s_i s_j) = 1 + \sum_{i<j} c_{ij} s_i s_j + \frac{1}{2} \sum_{i<j}\sum_{k<l} c_{ij}c_{kl} s_i s_j s_k s_l + \dots
\end{equation}
with a polynomial number of parameters $c_{ij}$ corresponds to an exponential number of parameters in the Hadamard-Walsh expansion.

However, we stress that neither expansion can \emph{a priori} offer a guarantee of
\emph{efficient} approximability: for a generic many-body
state, the coefficients $c_A$ need not necessarily decay quickly, and not all states are well approximated by keeping only
a few terms. In such cases, a faithful representation
requires an exponentially large number of coefficients, and the state is well represented only when \emph{all} the terms are taken into account. As we shall see below, this is strongly indicative of the RBM performance. In comparison to Sect.~\ref{subsec:phaseAmpUniformity}, we see that the uniform state is the special case of truncating the cumulant expansion with only the zeroth order term, valid when this term is much bigger in magnitude than all the others.

In what follows, we compare optimzed RBMs to exact wavefunctions through the convergence properties of this truncated cumulant expansion.

\paragraph{Relation to RBM performance}

Deep in the paramagnetic phase of the unrotated Hamiltonian, the ground state is well-approximated as a perturbation of the uniform superposition $\ket{+} = \ket{\uparrow,\uparrow,\dots}_x$. As argued in Ref.~\cite{borinApproximatingPowerMachinelearning2020}, RBMs can achieve excellent accuracy in this regime because they effectively reproduce high orders of the expansion of $e^{W}\ket{+}$, where $W$ is a diagonal matrix in the computational basis.
Expressing $W$ in powers of Pauli-$Z$ operators immediately leads to a cumulant expansion of the form in \cref{eq:clusterexp}.
These statements mostly refer to expressability of RBM's. Instead, we now investigate how the RBM's practical performance depends on the convergence properties of this expansion in the common \emph{learning} framework.

\begin{figure}
    \centering
    \includegraphics[width=1.0\linewidth]{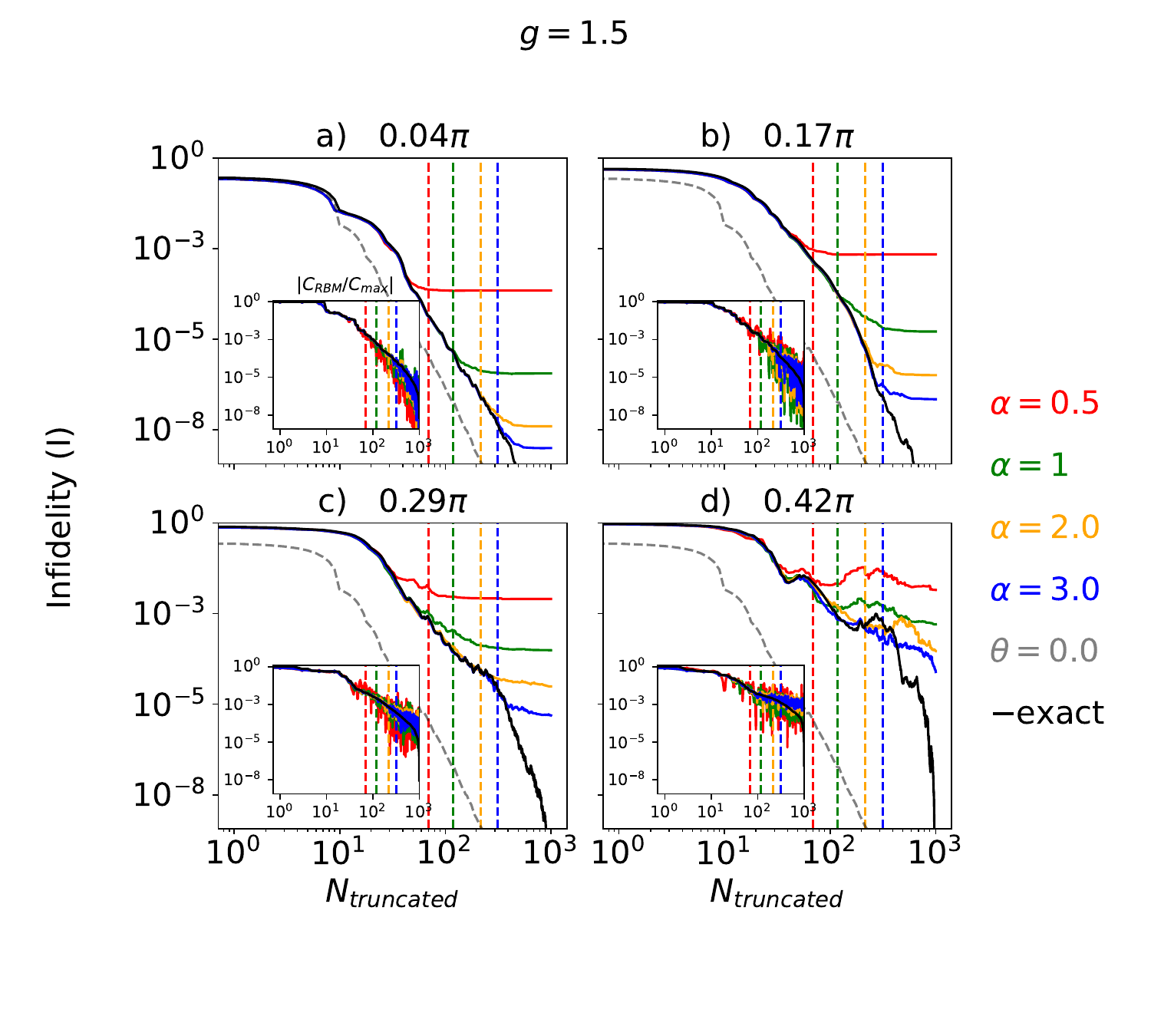}
    \caption{Infidelity over the number of coefficients of the truncated cumulant expansion based on \cref{eq:clusterexp}. The black line shows the truncation of the exact ground state. Truncations of RBMs with different hidden unit densities $\alpha$ are shown as colored lines. The vertical dashed lines indicate the number of variational parameters $N_\mathrm{var}$ of the corresponding RBM. The insets show the absolute values of the coefficients on which the truncation is based. Results are shown for different rotation angles $\theta$ at $g=1.5$ and $L=10$.}
    \label{fig:InfidelityG150L10}
\end{figure}

Figure~\ref{fig:InfidelityG150L10} compares the truncation of the exact ground state, $\Psi^{\left(\Ntrunc \right)}(s)$ (black line), with the truncation of the RBM state, $\Psi^{(\Ntrunc)}_{\mathrm{RBM}}(s)$ (colored lines corresponding to different $\alpha$), at $g=1.5$. 
We show the infidelity with respect to ground state $\ket{\Psi}$ as a function of number of kept coefficients $\Ntrunc$ for various angles $\theta$ and hidden-unit densities $\alpha$. This implies that the infidelity of $\Psi^{(\Ntrunc)}(s)$ must converge to $0$ as $\Ntrunc \to 2^L$, while the infidelity of $\Psi^{(\Ntrunc)}_{\mathrm{RBM}}(s)$ converges to a finite value, given by the total infidelity of the RBM state. We note that the infidelity can generally be non-monotonic as the incremental addition of parameters may break ground state symmetries.

Here, we find for all small $\Ntrunc$, truncated RBMs at various hidden unit densities agree well with the truncation of the exact state, indicating that the RBM accurately captures the dominant correlations.
Interestingly, as $\Ntrunc$ grows beyond the number of variational parameters $ N_\mathrm{var} = \alpha L^2 + \alpha L + L$, the infidelity of $\Psi^{(\Ntrunc)}_\mathrm{RBM}$ deviates from $\Psi^{(\Ntrunc)}$, demonstrating that the RBM predicts different values for less significant cluster coefficients. Moreover, in most cases the infidelity of the RBM state remains approximately constant beyond the threshold line $N= N_\mathrm{var}$. Thus in these cases, the \emph{full} RBM's performance is equivalent to that of a truncated cumulant expansion retaining only $N_\mathrm{var}$ coefficients. 
Surprisingly, this observation is seen to be robust with respect to $\theta$ and $g$ even at the critical point (see Appendix~\ref{appendix:cluster_expansion}).

A notable exception occurs precisely at $\theta=\pi/2$. Here, both expansions converge highly non-monotonically and the infidelity of the truncated RBM deviates sooner from the exact truncation, displaying highly oscillatory behavior at any finite truncation while ultimately recovering a low infidelity up until $10^{-8}$, despite also displaying significant errors for the coefficients themselves (see inset).

At $\theta=\pi/2$, the parity operator is diagonal in the computational basis. Because the parity operator commutes with the Hamiltonian and the ground state is unique, the ground state is guaranteed to have a well-defined parity of $+1$. As a result, odd-parity states $\hat{P}\ket{s}=(-1)\ket{s}$ do not contribute to the wavefunction, yielding $\Psi(s)=0$. 

\begin{figure}
    \centering
    \includegraphics[width=0.5\linewidth]{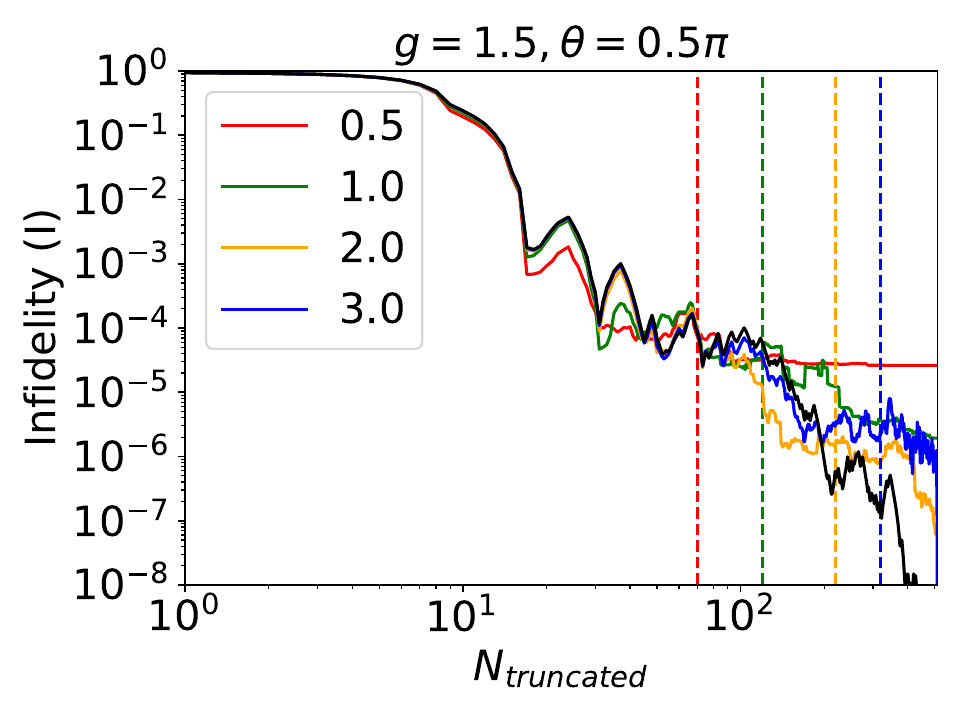}
    \caption{Infidelity of the truncated wavefunction with the exact state, analogous to Fig.~\ref{fig:InfidelityG150L10} in the reduced Hilbert space for $\theta=\pi/2$.}
    \label{fig:InfidelityG150L10_pi/2}
\end{figure}

Since the cluster expansion coefficients are related to the logarithm of the wavefunction, these zeros render the cluster expansion ill-defined across the full Hilbert space, requiring that the analysis be projected onto the correct parity sector. Figure \ref{fig:InfidelityG150L10_pi/2} displays the infidelity of the truncated expansion within this reduced Hilbert subspace. Even in this restricted sector, the expected trend holds, with the RBM accurately following the theoretical curve derived from the expansion.


    From this analyis we may also better understand the basis dependence of RBM performance. Evidently, for small $\theta$, the truncated cumulant expansion converges more rapidly compared to large $\theta$. As the RBM is seen to accurately capture only the dominant coefficients, this directly translates to a better performance at small $\theta$.

\section{Conclusions}\label{conclusions}
In this work, we have investigated the impact of the choice of basis for restricted Boltzmann machines, within the transverse field Ising model.
Employing basis rotations via a rotation of the Hamiltonian, we unraveled the complex interplay between the factors driving the performance of RBMs.
First, we demonstrate how the absence of a unique ground state can significantly hinder the RBM's ability to learn the true ground state, as it tends to converge to the simplest superposition of degenerate states.
Furthermore, we demonstrate that beyond spectral properties, the uniformity of \emph{both} phases and amplitudes of the ground state wavefunction in the computational basis is a crucial and often overlooked benefactor influencing RBM's performance.
In this context, we provide a new way to accurately predict the performance of RBM's based on the convergence properties of the cumulant expansion of the ground state wavefunction. To good accuracy, we find that the RBM's performance is equivalent to that of a cumulant expansion, which retains only the most significant (in terms of the magnitude of their relative coefficient) multi-body correlations up to the RBM's number of variational parameters.
As this expansion is highly basis-dependent, this finding provides unique insight into the identification of the correct basis for numerical computations. 

Based on our results, we recommend the following strategy, relying solely on exact diagonalization of small clusters, which may be employed \emph{a priori} to assess the suitability of RBM's for a given problem: (i) Identify the low-energy spectrum and check for near degeneracies. If present, it is likely that the RBM will converge to a superposition of the nearly degenerate states which may hinder accuracy drastically.
(ii) Compute the cumulant expansion of the ground state. Typically, an RBM with $N_\textrm{var}$ variational parameters is expected to perform similarly to a truncated cumulant expansion which retains only the $N_\textrm{var}$ terms with the largest coefficients.
(iii) Search for the basis transformation for which the truncated cumulant expansion converges most rapidly. 

We believe that these factors likely play a similar role in the performance of other network architectures applied to different models, as our considerations are based mainly on the complexity of representing a given probability distribution. However, questions regarding the generalizability of our findings are certainly relevant and left for future work.

\section{Acknowledgements}
We thank Federico Becca, Giuseppe Carleo, Rajat Kumar Panda, Riccardo Rende, and Luciano Loris Viteritti for helpful discussions. R.~V. also acknowledges discussions with Sven Benjamin Ko\v{z}i\'{c} on related aspects of NQS basis dependency. 
M.~D., R.~V., and N.~N.~ acknowledge funding from the European Research
Council (ERC) under the European Union’s Horizon ERC-2022-COG Grant with Project number 101087692. M.~D. was also supported by the EU-Flagship programme Pasquans2, by the PNRR MUR project PE0000023-NQSTI, and the PRIN programme (project CoQuS).

\bibliography{bib}

@misc{paul2025boundentanglementneuralquantum,
      title={Bound on entanglement in neural quantum states}, 
      author={Nisarga Paul},
      year={2025},
      eprint={2510.11797},
      archivePrefix={arXiv},
      primaryClass={quant-ph},
      url={https://arxiv.org/abs/2510.11797}, 
}

@article{PhysRevLett.134.076502,
  title = {Mott Transition and Volume Law Entanglement with Neural Quantum States},
  author = {Gauvin-Ndiaye, Chlo\'e and Tindall, Joseph and Moreno, Javier Robledo and Georges, Antoine},
  journal = {Phys. Rev. Lett.},
  volume = {134},
  issue = {7},
  pages = {076502},
  numpages = {8},
  year = {2025},
  month = {Feb},
  publisher = {American Physical Society},
  doi = {10.1103/PhysRevLett.134.076502},
  url = {https://link.aps.org/doi/10.1103/PhysRevLett.134.076502}
}

@article{PRXQuantum.3.010333,
  title = {Random Quantum Circuits Anticoncentrate in Log Depth},
  author = {Dalzell, Alexander M. and Hunter-Jones, Nicholas and Brand\~ao, Fernando G. S. L.},
  journal = {PRX Quantum},
  volume = {3},
  issue = {1},
  pages = {010333},
  numpages = {43},
  year = {2022},
  month = {Mar},
  publisher = {American Physical Society},
  doi = {10.1103/PRXQuantum.3.010333},
  url = {https://link.aps.org/doi/10.1103/PRXQuantum.3.010333}
}

@Article{mbeng2024quantum,
	title={{The quantum Ising chain for beginners}},
	author={Glen Bigan Mbeng and Angelo Russomanno and Giuseppe E. Santoro},
	journal={SciPost Phys. Lect. Notes},
	pages={82},
	year={2024},
	publisher={SciPost},
	doi={10.21468/SciPostPhysLectNotes.82},
	url={https://scipost.org/10.21468/SciPostPhysLectNotes.82},
}

@article{gu2025solving,
  title={Solving the Hubbard model with Neural Quantum States},
  author={Gu, Yuntian and Li, Wenrui and Lin, Heng and Zhan, Bo and Li, Ruichen and Huang, Yifei and He, Di and Wu, Yantao and Xiang, Tao and Qin, Mingpu and others},
  journal={arXiv preprint arXiv:2507.02644},
  year={2025}
}

@article{PhysRevB.97.035116,
  title = {Approximating quantum many-body wave functions using artificial neural networks},
  author = {Cai, Zi and Liu, Jinguo},
  journal = {Phys. Rev. B},
  volume = {97},
  issue = {3},
  pages = {035116},
  numpages = {8},
  year = {2018},
  month = {Jan},
  publisher = {American Physical Society},
  doi = {10.1103/PhysRevB.97.035116},
  url = {https://link.aps.org/doi/10.1103/PhysRevB.97.035116}
}

@article{PhysRevB.97.195136,
  title = {Chiral topological phases from artificial neural networks},
  author = {Kaubruegger, Raphael and Pastori, Lorenzo and Budich, Jan Carl},
  journal = {Phys. Rev. B},
  volume = {97},
  issue = {19},
  pages = {195136},
  numpages = {7},
  year = {2018},
  month = {May},
  publisher = {American Physical Society},
  doi = {10.1103/PhysRevB.97.195136},
  url = {https://link.aps.org/doi/10.1103/PhysRevB.97.195136}
}

@article{PhysRevB.96.205152,
  title = {Restricted Boltzmann machine learning for solving strongly correlated quantum systems},
  author = {Nomura, Yusuke and Darmawan, Andrew S. and Yamaji, Youhei and Imada, Masatoshi},
  journal = {Phys. Rev. B},
  volume = {96},
  issue = {20},
  pages = {205152},
  numpages = {8},
  year = {2017},
  month = {Nov},
  publisher = {American Physical Society},
  doi = {10.1103/PhysRevB.96.205152},
  url = {https://link.aps.org/doi/10.1103/PhysRevB.96.205152}
}

@article{Lange2025,
  title = {Simulating the Two-Dimensional $t\text{\ensuremath{-}}J$ Model at Finite Doping with Neural Quantum States},
  author = {Lange, Hannah and B\"ohler, Annika and Roth, Christopher and Bohrdt, Annabelle},
  journal = {Phys. Rev. Lett.},
  volume = {135},
  issue = {13},
  pages = {136504},
  numpages = {7},
  year = {2025},
  month = {Sep},
  publisher = {American Physical Society},
  doi = {10.1103/rc31-5hl9},
  url = {https://link.aps.org/doi/10.1103/rc31-5hl9}
}

@article{PhysRevB.103.165103,
  title = {Variational classical networks for dynamics in interacting quantum matter},
  author = {Verdel, Roberto and Schmitt, Markus and Huang, Yi-Ping and Karpov, Petr and Heyl, Markus},
  journal = {Phys. Rev. B},
  volume = {103},
  issue = {16},
  pages = {165103},
  numpages = {17},
  year = {2021},
  month = {Apr},
  publisher = {American Physical Society},
  doi = {10.1103/PhysRevB.103.165103},
  url ={https://link.aps.org/doi/10.1103/PhysRevB.103.165103}
}

@article{kuboGeneralizedCumulantExpansion1962,
  title = {Generalized Cumulant Expansion Method},
  author = {Kubo, Ryogo},
  year = 1962,
  journal = {Journal of the Physical Society of Japan},
  volume = {17},
  number = {7},
  eprint = {https://doi.org/10.1143/JPSJ.17.1100},
  pages = {1100--1120},
  doi = {10.1143/JPSJ.17.1100}
}

@article{bartlettCoupledclusterTheoryQuantum2007,
  title = {Coupled-cluster theory in quantum chemistry},
  author = {Bartlett, Rodney J. and Musia\l{}, Monika},
  journal = {Rev. Mod. Phys.},
  volume = {79},
  issue = {1},
  pages = {291--352},
  numpages = {0},
  year = {2007},
  month = {Feb},
  publisher = {American Physical Society},
  doi = {10.1103/RevModPhys.79.291},
  url = {https://link.aps.org/doi/10.1103/RevModPhys.79.291}
}

@inproceedings{akibaOptunaNextgenerationHyperparameter2019,
  title = {Optuna: A next-Generation Hyperparameter Optimization Framework},
  booktitle = {Proceedings of the 25th {{ACM SIGKDD}} International Conference on Knowledge Discovery and Data Mining},
  author = {Akiba, Takuya and Sano, Shotaro and Yanase, Toshihiko and Ohta, Takeru and Koyama, Masanori},
  year = 2019
}

@book{beccaQuantumMonteCarlo2017,
  title = {Quantum Monte Carlo Approaches for Correlated Systems},
  author = {Becca, Federico and Sorella, Sandro},
  year = 2017,
  publisher = {Cambridge University Press},
  address = {Cambridge}
}

@article{borinApproximatingPowerMachinelearning2020,
  title = {Approximating Power of Machine-Learning Ansatz for Quantum Many-Body States},
  author = {Borin, Artem and Abanin, Dmitry A.},
  year = 2020,
  month = may,
  journal = {Physical Review B},
  volume = {101},
  number = {19},
  pages = {195141},
  publisher = {American Physical Society},
  doi = {10.1103/PhysRevB.101.195141}
}

@article{bukovLearningGroundState2021,
  title = {Learning the Ground State of a Non-Stoquastic Quantum {{Hamiltonian}} in a Rugged Neural Network Landscape},
  author = {Bukov, Marin and Schmitt, Markus and Dupont, Maxime},
  year = 2021,
  month = jun,
  journal = {SciPost Physics},
  volume = {10},
  number = {6},
  pages = {147},
  issn = {2542-4653},
  doi = {10.21468/SciPostPhys.10.6.147},
  urldate = {2025-09-10},
  langid = {english}
}

@article{vicentiniNetKet3Machine2022,
  title = {{{NetKet}} 3: {{Machine}} Learning Toolbox for Many-Body Quantum Systems},
  author = {Vicentini, Filippo and Hofmann, Damian and Szab{\'o}, Attila and Wu, Dian and Roth, Christopher and Giuliani, Clemens and Pescia, Gabriel and Nys, Jannes and {Vargas-Calder{\'o}n}, Vladimir and Astrakhantsev, Nikita and Carleo, Giuseppe},
  year = 2022,
  journal = {SciPost Phys. Codebases},
  pages = {7},
  publisher = {SciPost},
  doi = {10.21468/SciPostPhysCodeb.7}
}

@article{carleoSolvingQuantumManybody2017,
  title = {Solving the Quantum Many-Body Problem with Artificial Neural Networks},
  author = {Carleo, Giuseppe and Troyer, Matthias},
  year = 2017,
  month = feb,
  journal = {Science},
  volume = {355},
  number = {6325},
  pages = {602--606},
  publisher = {American Association for the Advancement of Science (AAAS)},
  issn = {1095-9203},
  doi = {10.1126/science.aag2302}
}

@article{chooTwodimensionalFrustratedJ1J22019,
  title = {Two-Dimensional Frustrated {{J}}{$_1$}-{{J}}{$_2$} Model Studied with Neural Network Quantum States},
  author = {Choo, Kenny and Neupert, Titus and Carleo, Giuseppe},
  year = 2019,
  month = sep,
  journal = {Physical Review B},
  volume = {100},
  number = {12},
  pages = {125124},
  publisher = {American Physical Society},
  doi = {10.1103/PhysRevB.100.125124}
}

@article{denisAccurateNeuralQuantum2025,
  title = {Accurate Neural Quantum States for Interacting Lattice Bosons},
  author = {Denis, Zakari and Carleo, Giuseppe},
  year = 2025,
  month = jun,
  journal = {Quantum},
  volume = {9},
  pages = {1772},
  publisher = {Verein zur F\"orderung des Open Access Publizierens in den Quantenwissenschaften},
  issn = {2521-327X},
  doi = {10.22331/q-2025-06-17-1772}
}

@article{schmittQuantumDynamicsTransversefield2018,
  title = {Quantum Dynamics in Transverse-Field {{Ising}} Models from Classical Networks},
  author = {Schmitt, Markus and Heyl, Markus},
  year = 2018,
  journal = {SciPost Physics},
  volume = {4},
  pages = {013},
  publisher = {SciPost},
  doi = {10.21468/SciPostPhys.4.2.013}
}

@article{chenEmpoweringDeepNeural2024,
  title = {Empowering Deep Neural Quantum States through Efficient Optimization},
  author = {Chen, Ao and Heyl, Markus},
  year = 2024,
  journal = {Nature Physics},
  volume = {20},
  number = {9},
  pages = {1476--1481},
  doi = {10.1038/s41567-024-02566-1},
  da = {2024/09/01},
  isbn = {1745-2481},
  ty = {JOUR}
}

@article{chenEquivalenceRestrictedBoltzmann2018,
  title = {Equivalence of Restricted {{Boltzmann}} Machines and Tensor Network States},
  author = {Chen, Jing and Cheng, Song and Xie, Haidong and Wang, Lei and Xiang, Tao},
  year = 2018,
  month = feb,
  journal = {Physical Review B},
  volume = {97},
  number = {8},
  pages = {085104},
  publisher = {American Physical Society},
  doi = {10.1103/PhysRevB.97.085104}
}

@article{clarkUnifyingNeuralnetworkQuantum2018,
  title = {Unifying Neural-Network Quantum States and Correlator Product States via Tensor Networks},
  author = {Clark, Stephen R},
  year = 2018,
  month = feb,
  journal = {Journal of Physics A: Mathematical and Theoretical},
  volume = {51},
  number = {13},
  pages = {135301},
  publisher = {IOP Publishing},
  title+duplicate-1 = {Unifying neural-network quantum states and correlator product states via tensor networks}
}

@article{colluraDescriptivePowerNeuralNetworks2021,
  title = {On the Descriptive Power of {{Neural-Networks}} as Constrained {{Tensor Networks}} with Exponentially Large Bond Dimension},
  author = {Collura, Mario and Dell'Anna, Luca and Felser, Timo and Montangero, Simone},
  year = 2021,
  journal = {SciPost Phys. Core},
  volume = {4},
  pages = {001},
  publisher = {SciPost},
  doi = {10.21468/SciPostPhysCore.4.1.001}
}

@article{dashEfficiencyNeuralQuantum2024,
  title = {Efficiency of Neural Quantum States in Light of the Quantum Geometric Tensor},
  author = {Dash, Sidhartha and Gravina, Luca and Vicentini, Filippo and Ferrero, Michel and Georges, Antoine},
  year = 2024,
  journal = {arXiv preprint arXiv:2402.01565},
  eprint = {2402.01565},
  archiveprefix = {arXiv}
}

@article{dengQuantumEntanglementNeural2017,
  title = {Quantum Entanglement in Neural Network States},
  author = {Deng, Dong-Ling and Li, Xiaopeng and Das Sarma, S.},
  year = 2017,
  month = may,
  journal = {Physical Review X},
  volume = {7},
  number = {2},
  pages = {021021},
  publisher = {American Physical Society},
  doi = {10.1103/PhysRevX.7.021021}
}

@article{denisCommentCanNeural2025,
  title = {Comment on ``Can Neural Quantum States Learn Volume-Law Ground States?''},
  author = {Denis, Zakari and Sinibaldi, Alessandro and Carleo, Giuseppe},
  year = 2025,
  month = feb,
  journal = {Physical Review Letters},
  volume = {134},
  number = {7},
  pages = {079701},
  publisher = {American Physical Society},
  doi = {10.1103/PhysRevLett.134.079701}
}

@misc{doschlImportanceCorrelationsNeural2025,
  title = {Importance of Correlations for Neural Quantum States},
  author = {D{\"o}schl, Fabian and Bohrdt, Annabelle},
  year = 2025,
  eprint = {2508.14152},
  primaryclass = {quant-ph},
  archiveprefix = {arXiv}
}

@article{glasserNeuralnetworkQuantumStates2018,
  title = {Neural-Network Quantum States, String-Bond States, and Chiral Topological States},
  author = {Glasser, Ivan and Pancotti, Nicola and August, Moritz and Rodriguez, Ivan D. and Cirac, J. Ignacio},
  year = 2018,
  month = jan,
  journal = {Physical Review X},
  volume = {8},
  number = {1},
  pages = {011006},
  publisher = {American Physical Society},
  doi = {10.1103/PhysRevX.8.011006}
}

@article{huSpontaneousSymmetryBreaking2023,
  title = {Spontaneous Symmetry Breaking without Ground State Degeneracy in Generalized {{N-state}} Clock Model},
  author = {Hu, Yaozong and Watanabe, Haruki},
  year = 2023,
  month = may,
  journal = {Physical Review B},
  volume = {107},
  number = {19},
  pages = {195139},
  publisher = {American Physical Society},
  doi = {10.1103/PhysRevB.107.195139}
}

@article{langeArchitecturesApplicationsReview2024,
  title = {From Architectures to Applications: A Review of Neural Quantum States},
  author = {Lange, Hannah and {Van de Walle}, Anka and Abedinnia, Atiye and Bohrdt, Annabelle},
  year = 2024,
  month = sep,
  journal = {Quantum Science and Technology},
  volume = {9},
  number = {4},
  pages = {040501},
  publisher = {IOP Publishing},
  doi = {10.1088/2058-9565/ad7168}
}

@article{lerouxRepresentationalPowerRestricted2008,
  title = {Representational Power of Restricted Boltzmann Machines and Deep Belief Networks},
  author = {Le Roux, Nicolas and Bengio, Yoshua},
  year = 2008,
  month = jun,
  journal = {Neural Computation},
  volume = {20},
  number = {6},
  eprint = {https://direct.mit.edu/neco/article-pdf/20/6/1631/817339/neco.2008.04-07-510.pdf},
  pages = {1631--1649},
  issn = {0899-7667},
  doi = {10.1162/neco.2008.04-07-510}
}

@misc{machaczekNeuralQuantumState2025,
  title = {Neural {{Quantum State Study}} of {{Fracton Models}}},
  author = {Machaczek, Marc and Pollet, Lode and Liu, Ke},
  year = 2025,
  month = feb,
  number = {arXiv:2406.11677},
  eprint = {2406.11677},
  primaryclass = {quant-ph},
  publisher = {arXiv},
  doi = {10.48550/arXiv.2406.11677},
  urldate = {2025-02-18},
  archiveprefix = {arXiv}
}

@article{melkoRestrictedBoltzmannMachines2019,
  title = {Restricted {{Boltzmann}} Machines in Quantum Physics},
  author = {Melko, Roger G. and Carleo, Giuseppe and Carrasquilla, Juan and Cirac, J. Ignacio},
  year = 2019,
  journal = {Nature Physics},
  volume = {15},
  number = {9},
  pages = {887--892},
  doi = {10.1038/s41567-019-0545-1},
  da = {2019/09/01},
  isbn = {1745-2481},
  ty = {JOUR}
}

@article{nomuraRestrictedBoltzmannMachine2017,
  title = {Restricted {{Boltzmann}} Machine Learning for Solving Strongly Correlated Quantum Systems},
  author = {Nomura, Yusuke and Darmawan, Andrew S. and Yamaji, Youhei and Imada, Masatoshi},
  year = 2017,
  month = nov,
  journal = {Physical Review B},
  volume = {96},
  number = {20},
  pages = {205152},
  publisher = {American Physical Society},
  doi = {10.1103/PhysRevB.96.205152}
}

@article{ouImprovingNeuralNetwork2025,
  title = {Improving Neural Network Performance for Solving Quantum Sign Structure},
  author = {Ou, Xiaowei and Huang, Tianshu and Ozoli{\c n}{\v s}, Vidvuds},
  year = 2025,
  month = oct,
  journal = {Physical Review B},
  volume = {112},
  number = {16},
  pages = {165122},
  publisher = {American Physical Society},
  doi = {10.1103/fqxr-r8vw}
}

@misc{chenNeuralNetworkaugmentedPfaffian2025,
  title = {Neural Network-Augmented Pfaffian Wave-Functions for Scalable Simulations of Interacting Fermions},
  author = {Chen, Ao and Wan, Zhou-Quan and Sengupta, Anirvan and Georges, Antoine and Roth, Christopher},
  year = 2025,
  eprint = {2507.10705},
  primaryclass = {cond-mat.str-el},
  archiveprefix = {arXiv},
  journal = {arXiv}
}

@article{viterittiAccuracyRestrictedBoltzmann2022,
  title = {Accuracy of Restricted {{Boltzmann}} Machines for the One-Dimensional {{J}}{$_1$}-{{J}}{$_2$} {{Heisenberg}} Model},
  author = {Viteritti, Luciano Loris and Ferrari, Francesco and Becca, Federico},
  year = 2022,
  journal = {SciPost Physics},
  volume = {12},
  pages = {166},
  publisher = {SciPost},
  doi = {10.21468/SciPostPhys.12.5.166}
}

@article{PhysRevLett.130.236401,
  title = {Transformer Variational Wave Functions for Frustrated Quantum Spin Systems},
  author = {Viteritti, Luciano Loris and Rende, Riccardo and Becca, Federico},
  journal = {Phys. Rev. Lett.},
  volume = {130},
  issue = {23},
  pages = {236401},
  numpages = {6},
  year = {2023},
  month = {Jun},
  publisher = {American Physical Society},
  doi = {10.1103/PhysRevLett.130.236401},
  url = {https://link.aps.org/doi/10.1103/PhysRevLett.130.236401}
}

@article{chenNeuralNetworkEvolution2022,
 title = {Neural network evolution strategy for solving quantum sign structures},
  author = {Chen, Ao and Choo, Kenny and Astrakhantsev, Nikita and Neupert, Titus},
  journal = {Phys. Rev. Res.},
  volume = {4},
  issue = {2},
  pages = {L022026},
  numpages = {6},
  year = {2022},
  month = {May},
  publisher = {American Physical Society},
  doi = {10.1103/PhysRevResearch.4.L022026},
  url = {https://link.aps.org/doi/10.1103/PhysRevResearch.4.L022026}
}

@article{viterittiTransformerWaveFunction2025,
  title = {Transformer Wave Function for Two Dimensional Frustrated Magnets: {{Emergence}} of a Spin-Liquid Phase in the {{Shastry-Sutherland}} Model},
  author = {Viteritti, Luciano Loris and Rende, Riccardo and Parola, Alberto and Goldt, Sebastian and Becca, Federico},
  year = 2025,
  month = apr,
  journal = {Physical Review B},
  volume = {111},
  number = {13},
  pages = {134411},
  publisher = {American Physical Society},
  doi = {10.1103/PhysRevB.111.134411}
}

@article{parkExpressivePowerComplexvalued2022,
  title = {Expressive Power of Complex-Valued Restricted {{Boltzmann}} Machines for Solving Nonstoquastic Hamiltonians},
  author = {Park, Chae-Yeun and Kastoryano, Michael J.},
  year = 2022,
  month = oct,
  journal = {Physical Review B},
  volume = {106},
  number = {13},
  pages = {134437},
  publisher = {American Physical Society},
  doi = {10.1103/PhysRevB.106.134437}
}

@article{parkGeometryLearningNeural2020,
  title = {Geometry of Learning Neural Quantum States},
  author = {Park, Chae-Yeun and Kastoryano, Michael J.},
  date = {2020-05},
  year = {2020},
  journal = {Phys. Rev. Res.},
  volume = {2},
  number = {2},
  pages = {023232},
  publisher = {American Physical Society},
  doi = {10.1103/PhysRevResearch.2.023232},
  url = {https://link.aps.org/doi/10.1103/PhysRevResearch.2.023232},
  pagetotal = {16}
}

@article{szaboNeuralNetworkWave2020,
  title = {Neural Network Wave Functions and the Sign Problem},
  author = {Szabó, Attila and Castelnovo, Claudio},
  date = {2020-07},
  year = {2020},
  journal = {Phys. Rev. Res.},
  volume = {2},
  number = {3},
  pages = {033075},
  publisher = {American Physical Society},
  doi = {10.1103/PhysRevResearch.2.033075},
  url = {https://link.aps.org/doi/10.1103/PhysRevResearch.2.033075},
  pagetotal = {12}
}

@article{triguerosSimplicityMeanfieldTheories2024,
  title = {Simplicity of Mean-Field Theories in Neural Quantum States},
  author = {Trigueros, Fabian Ballar and Mendes-Santos, Tiago and Heyl, Markus},
  date = {2024-06},
  year = {2024},
  journal = {Phys. Rev. Res.},
  volume = {6},
  number = {2},
  pages = {023261},
  publisher = {American Physical Society},
  doi = {10.1103/PhysRevResearch.6.023261},
  url = {https://link.aps.org/doi/10.1103/PhysRevResearch.6.023261},
  pagetotal = {9}
}

@article{peiCompactNeuralnetworkQuantum2021,
  title = {Compact Neural-Network Quantum State Representations of {{Jastrow}} and Stabilizer States},
  author = {Pei, Michael Y and Clark, Stephen R},
  year = 2021,
  month = sep,
  journal = {Journal of Physics A: Mathematical and Theoretical},
  volume = {54},
  number = {40},
  pages = {405304},
  publisher = {IOP Publishing},
  issn = {1751-8121},
  doi = {10.1088/1751-8121/ac1f3d}
}

@book{sachdevQuantumPhaseTransitions2011,
  title = {Quantum Phase Transitions},
  author = {Sachdev, Subir},
  year = 2011,
  edition = {2},
  publisher = {Cambridge University Press, Cambridge},
  address = {Cambridge},
  doi = {10.1017/CBO9780511973765}
}

@article{schurovLearningComplexityManybody2025,
  title = {Learning Complexity of Many-Body Quantum Sign Structures through the Lens of Boolean Fourier Analysis},
  author = {Schurov, Ilya and Kravchenko, Anna and Katsnelson, Mikhail I and Bagrov, Andrey A and Westerhout, Tom},
  year = 2025,
  journal = {arXiv preprint arXiv:2508.09870},
  eprint = {2508.09870},
  archiveprefix = {arXiv}
}

@article{sharirNeuralTensorContractions2022,
  title = {Neural Tensor Contractions and the Expressive Power of Deep Neural Quantum States},
  author = {Sharir, Or and Shashua, Amnon and Carleo, Giuseppe},
  year = 2022,
  month = nov,
  journal = {Physical Review B},
  volume = {106},
  number = {20},
  pages = {205136},
  publisher = {American Physical Society},
  doi = {10.1103/PhysRevB.106.205136}
}

@misc{sinibaldiNetket_fidelityPackage2023,
  title = {Netket\_fidelity {{Package}}},
  author = {Sinibaldi, Alessandro and Vicentini, Filippo},
  year = 2023,
  doi = {10.5281/zenodo.8344170}
}

@article{valentiCorrelationenhancedNeuralNetworks2022,
  title = {Correlation-enhanced neural networks as interpretable variational quantum states},
  author = {Valenti, Agnes and Greplova, Eliska and Lindner, Netanel H. and Huber, Sebastian D.},
  journal = {Phys. Rev. Res.},
  volume = {4},
  issue = {1},
  pages = {L012010},
  numpages = {7},
  year = {2022},
  month = {Jan},
  publisher = {American Physical Society},
  doi = {10.1103/PhysRevResearch.4.L012010},
  url = {https://link.aps.org/doi/10.1103/PhysRevResearch.4.L012010}
}

@misc{yangWhenCanClassical2024,
  title = {When Can Classical Neural Networks Represent Quantum States?},
  author = {Yang, Tai-Hsuan and Soleimanifar, Mehdi and Bergamaschi, Thiago and Preskill, John},
  year = 2024,
  eprint = {2410.23152},
  primaryclass = {quant-ph},
  archiveprefix = {arXiv}
}

@article{calabreseEntanglementEntropyConformal2009,
  title = {Entanglement Entropy and Conformal Field Theory},
  author = {Calabrese, Pasquale and Cardy, John},
  year = 2009,
  month = dec,
  journal = {Journal of Physics A: Mathematical and Theoretical},
  volume = {42},
  number = {50},
  pages = {504005},
  doi = {10.1088/1751-8113/42/50/504005}
}

@article{jastrowManybodyProblemStrong1955,
  title = {Many-Body Problem with Strong Forces},
  author = {Jastrow, Robert},
  year = 1955,
  month = jun,
  journal = {Physical Review},
  volume = {98},
  number = {5},
  pages = {1479--1484},
  publisher = {American Physical Society},
  doi = {10.1103/PhysRev.98.1479}
}

@article{seeding_kaneko,
  title = {Seeding neural network quantum states with tensor network states},
  author = {Kaneko, Ryui and Goto, Shimpei},
  journal = {Phys. Rev. B},
  volume = {112},
  issue = {15},
  pages = {155163},
  numpages = {15},
  year = {2025},
  month = {Oct},
  publisher = {American Physical Society},
  doi = {10.1103/3rq4-2m5k},
  url = {https://link.aps.org/doi/10.1103/3rq4-2m5k}
}

@article{malyshev2024neural,
  title={Neural Quantum States and Peaked Molecular Wave Functions: Curse or Blessing?},
  author={Malyshev, Aleksei and Schmitt, Markus and Lvovsky, AI},
  journal={arXiv preprint arXiv:2408.07625},
  year={2024}
}

@article{kovzic2025exploring,
  title={Exploring the Effect of Basis Rotation on NQS Performance},
  author={Ko{\v{z}}i{\'c}, Sven Benjamin and Zlati{\'c}, Vinko and Franchini, Fabio and Giampaolo, Salvatore Marco},
  journal={arXiv preprint arXiv:2512.17893},
  year={2025}
}

@article{Sobral_2025,
   title={Physics-informed transformers for electronic quantum states},
   volume={16},
   ISSN={2041-1723},
   url={http://dx.doi.org/10.1038/s41467-025-66844-z},
   DOI={10.1038/s41467-025-66844-z},
   number={1},
   journal={Nature Communications},
   publisher={Springer Science and Business Media LLC},
   author={Sobral, João Augusto and Perle, Michael and Scheurer, Mathias S.},
   year={2025},
   month=nov }

\begin{appendices}

\section{Details of the numerical simulations}
 \label{sec:netket}
 
The RBM simulations were performed using \textsc{NetKet}~\cite{vicentiniNetKet3Machine2022,sinibaldiNetket_fidelityPackage2023}. For the stochastic reconfiguration procedure, we conducted a hyperparameter search with \textsc{Optuna}~\cite{akibaOptunaNextgenerationHyperparameter2019}, performing $100$ random trials with learning rates sampled uniformly within the range $[10^{-5}, 10^{-1}]$. Since we have considered small system sizes, we compute all the relevant expectation values by summing over all basis states. In addition, for each set of physical parameters, we consider $10$ independent realizations of the RBM calculations whose parameters were randomly initialized using different seeds. The number of iterations in the optimization loop was set to $N_\mathrm{iter}=2000$, ensuring convergence in our simulations. 

\paragraph{Walsh-Hadamard transform}
\label{sec:Hadamard}
Closely related to the cumulant expansion is the Walsh-\\Hadamard transformation, the binary analogue of Fourier transform, commonly used in digital signal processing and in image compression.  
It is defined via expanding $\Psi$ in the form
\begin{equation}
    \Psi(s)=\sum_{A}C_A s_A,
\end{equation}
The coefficients $C_A$ can be obtained most efficiently in $\mathcal{O}(M\log M)$ complexity using the fast Walsh-Hadamard transform (FWHT) algorithm, where $M=2^L$. The transformation is its own inverse (up to an overall prefactor). As a quantum circuit, this corresponds to transforming the wavefunction by applying a Hadamard gate on each qubit.

Recent studies using such a decomposition have been carried out in the context of NQS in Refs.~\cite{doschlImportanceCorrelationsNeural2025, schurovLearningComplexityManybody2025}, and earlier in related contexts in Refs.~\cite{glasserNeuralnetworkQuantumStates2018,valentiCorrelationenhancedNeuralNetworks2022,borinApproximatingPowerMachinelearning2020}. The cumulant expansion in Eq.~\eqref{eq:clusterexp} can then be described as the Walsh-Hadamard transform of $\log\Psi$. However, one must note that even truncating the cumulant expansion at order 2, resulting in a Jastrow wave function, contains all orders of the Walsh-Hadamard decomposition. 

\paragraph{Relation to the cumulant expansion}
For general complex coefficients $C_A$, the set of boolean monomials forms a complete basis.
There are $2^N$ different $C_A$ terms in this sum, $2^N$ components of the wave function, indicating a one-to-one mapping between the two. The coefficients $c_A$ of the cumulant expansion are related to the $C_A$, albeit by a non-local mapping. In both cases, for a normalized wavefunction $\Psi$, there are only $2^N -1$ independent terms, with the zeroth order coefficient representing an overall normalization and an overall global phase.
Though truncations of both the cumulant and Walsh-Hadamard transforms offer a good way to compress a quantum state, we argue that the cumulant expansion more naturally captures the nature of correlations present in the state. 

\section{Derivation of Marshall sign rule}
To observe the sign structure induced, let us take a look at the quantity $\braket{s}{i\hat{\sigma}^y_j|\Psi_{\omega}}$.
\begin{equation}
    \braket{s}{i\hat{\sigma}^y_j|\Psi_{\omega}}= 
    \begin{cases}
        - \Psi_\omega(s') & \text{if } s_j=-1 \\
        + \Psi_\omega(s') & \text{if } s_j=1
    \end{cases}
    ,\quad \text{with} \; s'=(s_1,...,-s_j,...,s_N)
\label{eq:stoquastic_hams}
\end{equation}

Thus, a full rotation $e^{i\frac{\pi}{2}\sum_j \hat{\sigma}^y_j}$, is equivalent to a phase factor $e^{i\pi N_{\downarrow}}$, where the number of down spins $N_{\downarrow}=\frac{1}{2}\sum_j(1-s_j)$.  This replacement corresponds to a variation of a Marshall sign rule, as detailed in Ref. \cite{parkExpressivePowerComplexvalued2022}. 
\section{Exact application of a \texorpdfstring{$\pi$}{Pi} rotation to the RBM.}
\label{sec:pi_rotation_RBM}
Here, following Ref.~\cite{peiCompactNeuralnetworkQuantum2021}, we demonstrate, how a rotation of $\theta=\pi$ can be incorporated exactly using a modification of the RBM's variational parameters.
Starting from equation \ref{eq:stoquastic_hams}, considering $s'=(s_1,...,-s_j,...,s_N)$
and following the RBM definition, we observe that $\braket{s}{i\sigma^y_j|\Psi_{\omega}}=e^{i\frac{\pi}{2}(1-s_j)}\Psi_\omega(s')=\Psi_{\omega'}(s)$ with:
\begin{align*}
    a'_k &=
    \begin{cases}
        -\left(a_k + i\frac{\pi}{2}\right) & \text{for } k = j \\
        a_k & \text{otherwise}
    \end{cases}, \quad
    W'_{kl} =
    \begin{cases}
        -W_{kl} & \text{for } k = j \\
        W_{kl} & \text{otherwise}
    \end{cases},\quad
    b'_k = b_k
\end{align*}
Therefore, the application of a $\pi$ rotation maintains the RBM representation of the state.

\section{Cumulant expansion of the RBM state}
\label{appendix:cluster_expansion}

\begin{figure}[H]
    \centering
    \includegraphics[width=1.0\linewidth]{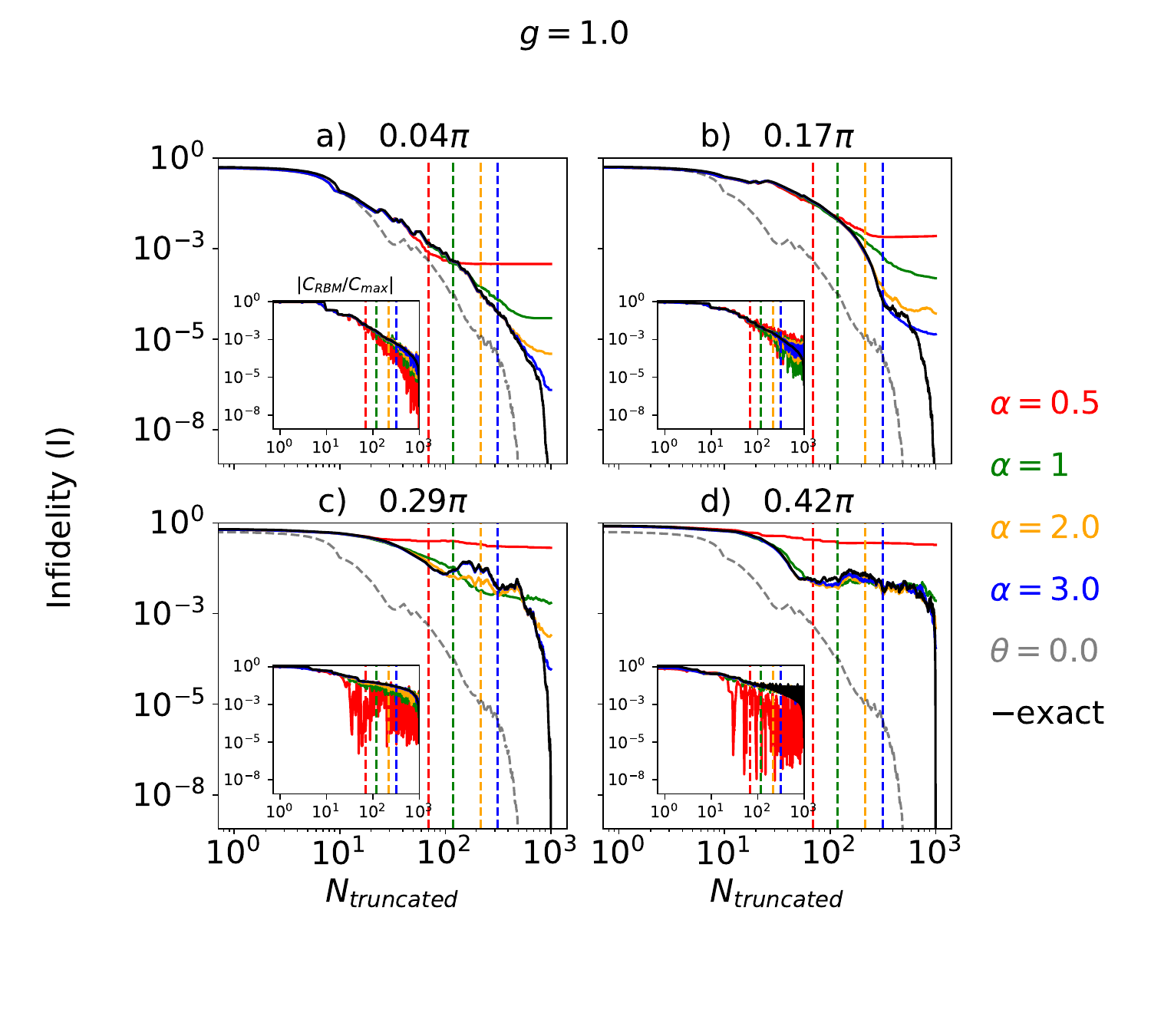}
    \caption{Infidelity over the number of coefficients of the truncated cumulant expansion. Refer to \cref{fig:InfidelityG150L10} for details.}
    \label{fig:InfidelityG100L10}
\end{figure}

Figure \ref{fig:InfidelityG100L10} shows the infidelity of the truncated cumulant expansion for $g = 1.0$. The general observation of the RBM's performance being limited to the first $N_\textrm{var}$ terms of the truncated cumulant expansion in the paramagnetic case (\cref{fig:InfidelityG150L10}) persists for most cases. A marked exception occurs at $\theta = 0.04\pi$, where the RBM follows the exact expansion even beyond the threshold indicated by the vertical line, marking the number of parameters of the RBM. This result suggests that the RBM is able to learn a more efficient encoding of the ground state. While the precise nature of this encoding is not known, it is possible that the RBM is able to learn certain symmetries of the coefficients such as $c_1 = c_L$.
In the other limit, $\theta=\pi/2$ the RBM also displays deviating behavior, predicting nearly constant amplitudes for the cumulant expansion coefficients. Despite the visibly large inaccuracy in the coefficients a low infidelity is recovered once all of them are taken into account.

\begin{figure}[H]
    \centering
    \includegraphics[width=0.65\linewidth]{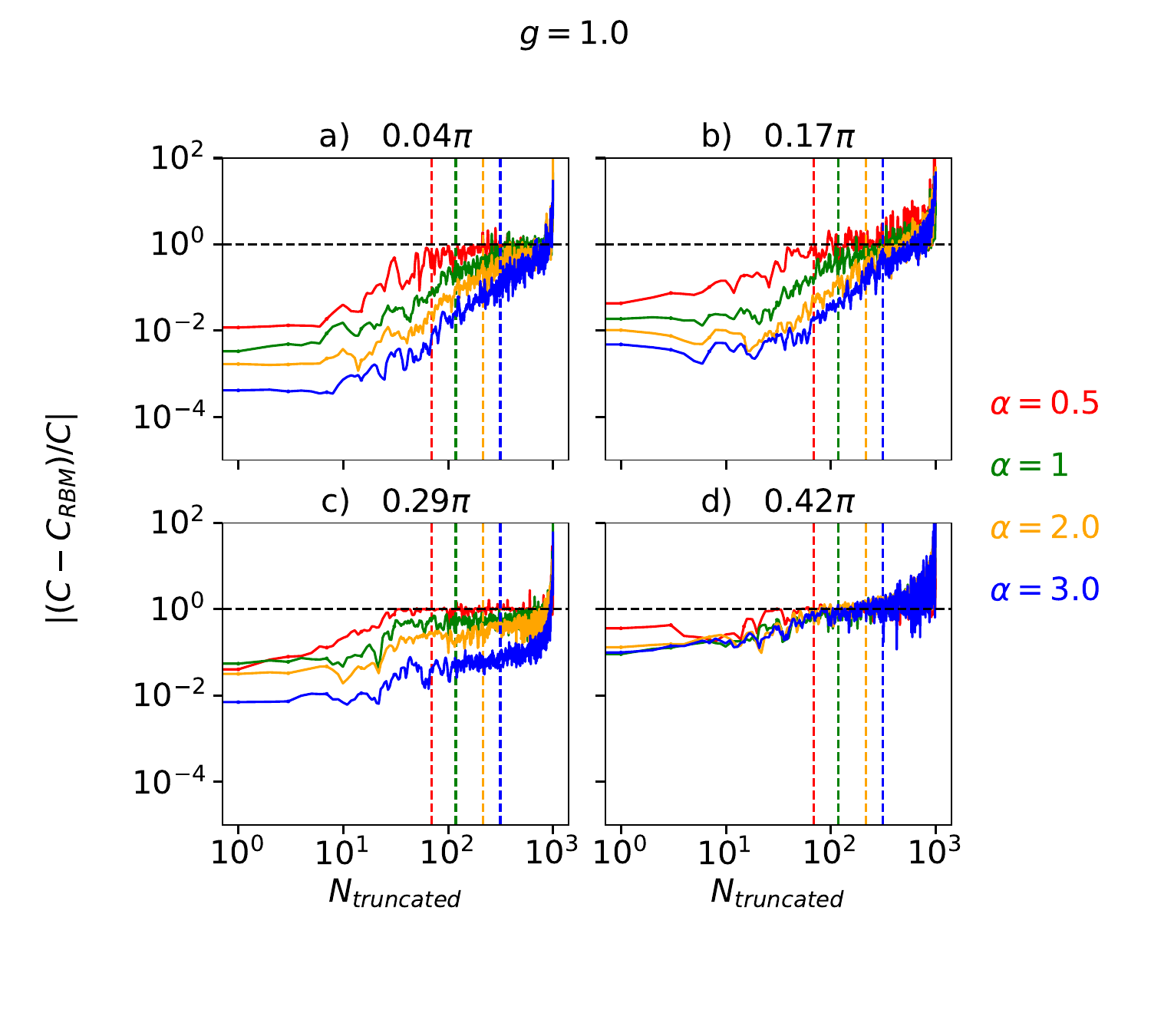}
    \caption{Relative error of the coefficients in the truncated cumulant expansion. The vertical dashed lines indicate the number of parameters, the black horizontal dashed line corresponds to a relative error of $1$.}
    \label{fig:RelerrG100L10}
\end{figure}

Figure~\ref{fig:RelerrG100L10} shows the relative errors for each coefficient in the truncated cumulant expansion. The black line marks the point where the relative error exceeds a value of $1$, corresponding to the RBM's prediction of nearly vanishing coefficients. For all angles except $\theta = 0.5\pi$, the relative error for coefficients  remains smaller than one below the threshold $\Ntrunc = N_\textrm{var}$. However, for $\theta = 0.5\pi$ this trend breaks down. In particular, for $\alpha > 0.5$ the relative error crosses $1$ much earlier.

\begin{figure}[H]
    \centering
    \includegraphics[width=0.65\linewidth]{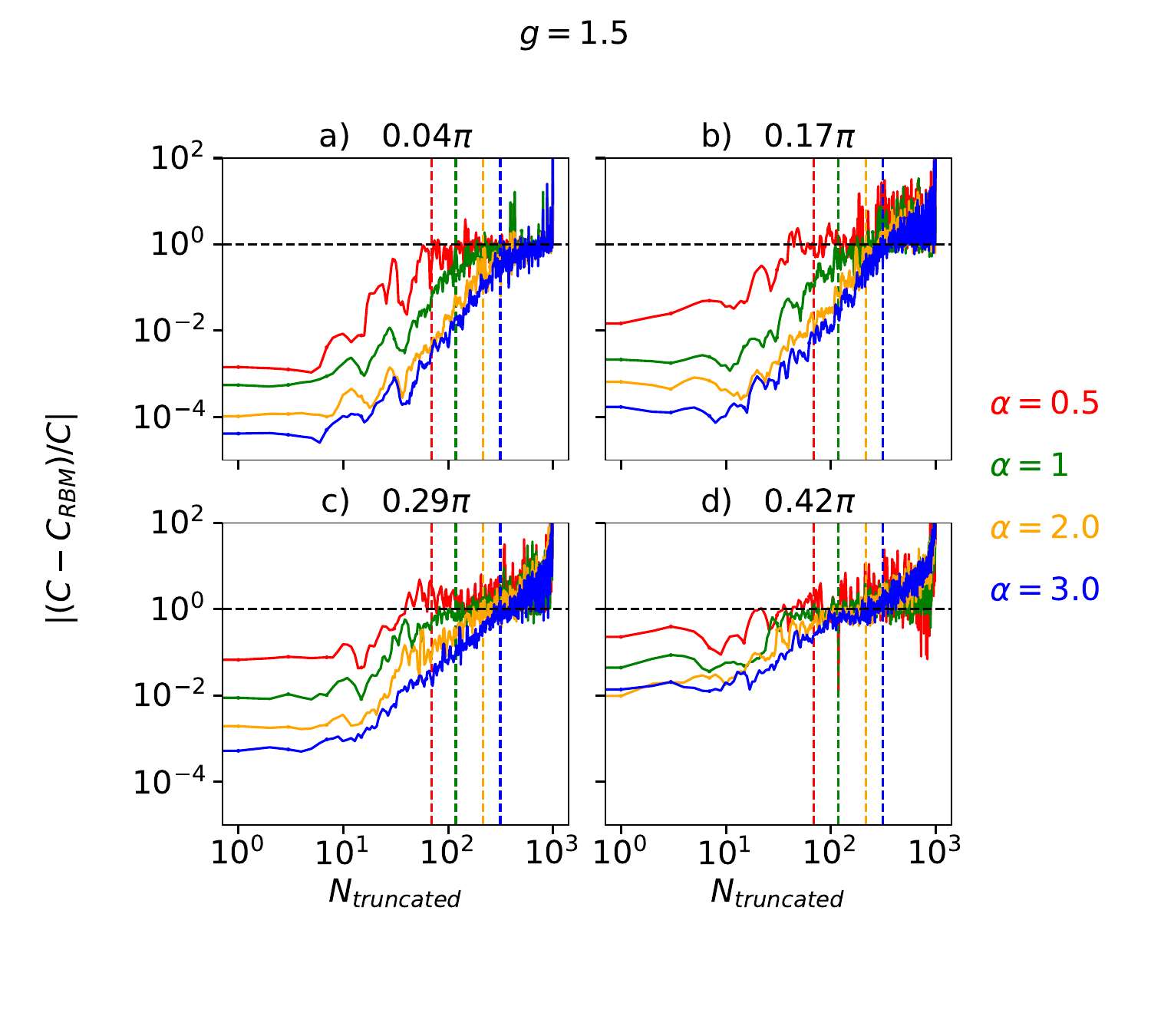}
    \caption{Relative error of the coefficients in the truncated cumulant expansion. The vertical dashed lines indicate the number of parameters, the black horizontal dashed line corresponds to a relative error of $1$.}
    \label{fig:RelerrG150L10}
\end{figure}

\section{Effect of increasing system size}
\begin{figure}[H]
    \centering
    \includegraphics[width=1.0\linewidth]{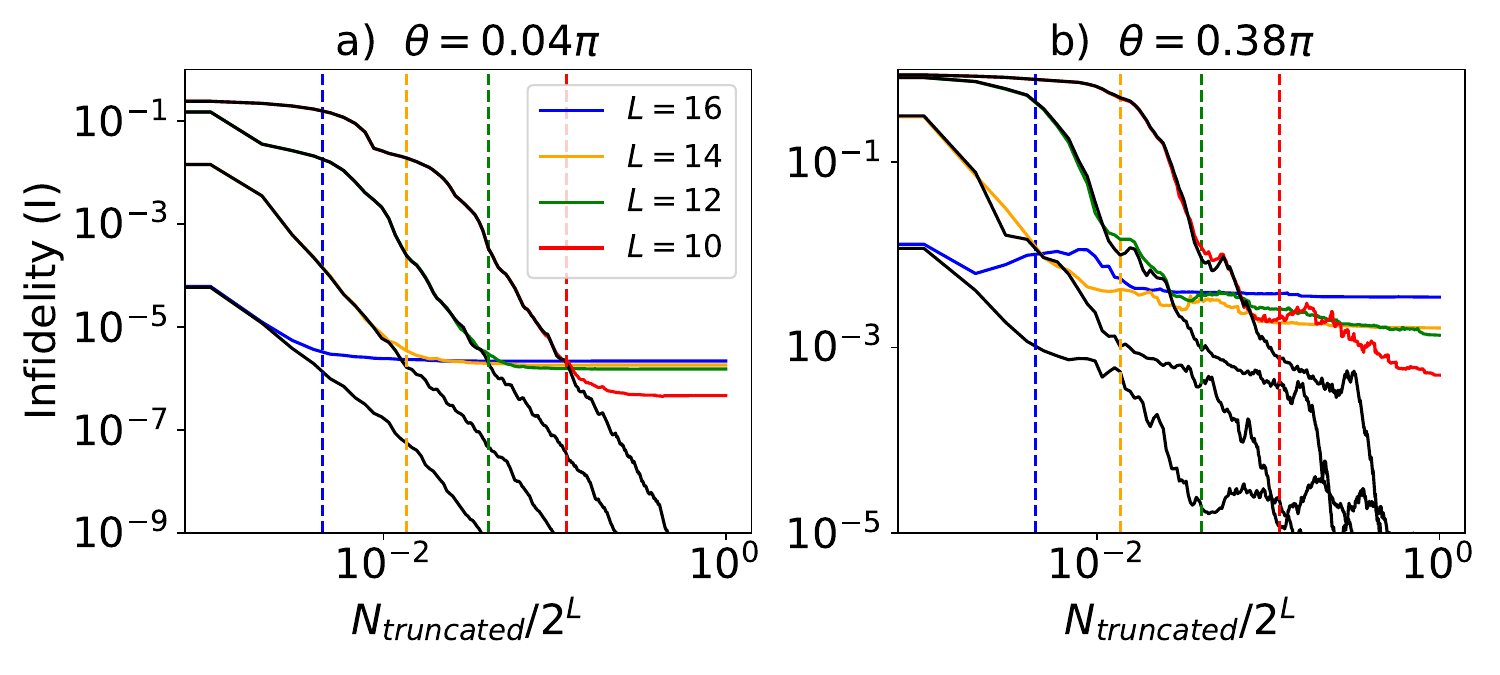}
    \caption{Infidelity curves vs the number of coefficients kept in the cluster expansion for $g=1.5$, the angles $\theta=0$ and $0.38\pi$ evaluated for different system sizes: $L=10,12,14$ and $16$.}
    \label{fig:Infidelity_vs_L}
\end{figure}

To investigate how the behavior observed in Figure~\ref{fig:InfidelityG150L10} scales with increasing the system size, we analyze system lengths  $L=10,12,14$ and $16$ for $g=1.5$ and two different angles. We choose two representative angles $\theta=0.04\pi$ and $\theta=0.38\pi$, and always keep the hidden unit density $\alpha=1$. Figure~\ref{fig:Infidelity_vs_L} shows that the main findings of this article -- that a rapidly convergent cumulant expansion is indicative of better performance of the RBM -- is robust to increasing system size.
For longer chains, the number of variational parameters in the neural network increases polynomially, whereas the total number of Hilbert space states increases exponentially. This explains the order of the dashed vertical lines in Fig.~\ref{fig:Infidelity_vs_L}, and hints at why the RBM performance is much better for smaller system sizes.

\end{appendices}
\end{document}